\documentclass[aps,a4paper]{revtex4}
\pdfoutput=1
\usepackage{epsfig}
\usepackage{graphicx}
\usepackage{subfig}
\usepackage{amsmath,amssymb,color}
\usepackage[english]{babel}
\usepackage{epstopdf}
\usepackage{xcolor}
\usepackage{dsfont}
\usepackage{amsthm}
\parskip=\medskipamount

\newcommand{\yury}[1]{{\color{blue}#1}}

\newcommand{\eq}[1]{(\ref{#1})}
\newcommand{\fig}[1]{Fig.~\ref{#1}}

\newcommand{\be}{\begin{equation}}
\newcommand{\ee}{\end{equation}}

\newcommand{\bS}{\mathbf{S}}
\newcommand{\bE}{\mathbf{E}}
\newcommand{\bT}{\mathbf{T}}

\newcommand{\bA}{\mathbf{A}}
\newcommand{\calB}{\mathcal{B}}
\newcommand{\calT}{\mathcal{T}}
\newcommand{\calS}{\mathcal{S}}

\newcommand{\tree}{\text{tree}}

\newcommand{\bd}{\mathbf{d}}

\newcommand{\NN}{\mathbb{N}}
\newcommand{\RR}{\mathbb{R}}
\newcommand{\II}{\mathds{1}}
\newcommand{\PP}{\mathbb{P}}
\newcommand{\EE}{\mathbb{E}}
\renewcommand{\mod}{\text{mod }}
\newcommand{\tM}{\text{M}}
\newcommand\scalemath[2]{\scalebox{#1}{\mbox{\ensuremath{\displaystyle #2}}}}

\theoremstyle{definition}

\begin{document}

\title{Peculiar spectral statistics of ensembles of trees and star-like graphs}

\author{V. Kovaleva$^{1}$, Yu. Maximov$^{1,2}$, S. Nechaev$^{3,4}$, and O. Valba$^{5,6}$}

\affiliation{$^{1}$ Center of Energy Systems, Skolkovo Institute of Science and Technology, Russia \\ $^{2}$ Center for Nonlinear Studies and Theoretical Division T-4, Los Alamos National Laboratory, Los Alamos, NM 87545, USA \\ $^{3}$CNRS/Independent University of Moscow, Poncelet Laboratory, Moscow, Russia \\ $^{4}$P.N. Lebedev Physical Institute RAS, Moscow, Russia \\
$^{5}$Department of Applied Mathematics, National Research University Higher School of Economics \\
$^{6}$N.N. Semenov Institute of Chemical Physics of the Russian Academy of Sciences}

\begin{abstract}
In this paper we investigate the eigenvalue statistics of exponentially weighted ensembles of full binary trees and $p$-branching star graphs. We show that spectral densities of corresponding
adjacency matrices demonstrate peculiar ultrametric structure inherent to sparse systems. In
particular, the tails of the distribution for binary trees share the ``Lifshitz singularity''
emerging in the one-dimensional localization, while the spectral statistics of $p$-branching
star-like graphs is less universal, being strongly dependent on $p$. The hierarchical structure of spectra of adjacency matrices is interpreted as sets of resonance frequencies, that emerge in ensembles of fully branched tree-like systems, known as dendrimers. However, the relaxational spectrum is not determined by the cluster topology, but has rather the number-theoretic origin, reflecting the peculiarities of the rare-event statistics typical for one-dimensional systems with a quenched structural disorder. The similarity of spectral densities of an individual dendrimer and of ensemble of linear chains with exponential distribution in lengths, demonstrates that dendrimers could be served as simple disorder-less toy models of one-dimensional systems with quenched disorder.

\end{abstract}

\maketitle

\section{Introduction}
One can gain the information about topological and statistical properties of polymers in solutions
by measuring their relaxation spectra \cite{brouwer2011spectra}. A rough model of an individual
polymer molecule of any topology is a set of monomers (atoms) connected by elastic strings. If
deformations of strings are small, the response of the molecule on external excitation is harmonic
according to the Hooke's law. The relaxation modes are basically determined by the so-called
Laplacian matrix (defined below) of the molecule.

\begin{figure}[ht]
\centerline{\includegraphics[width=7cm]{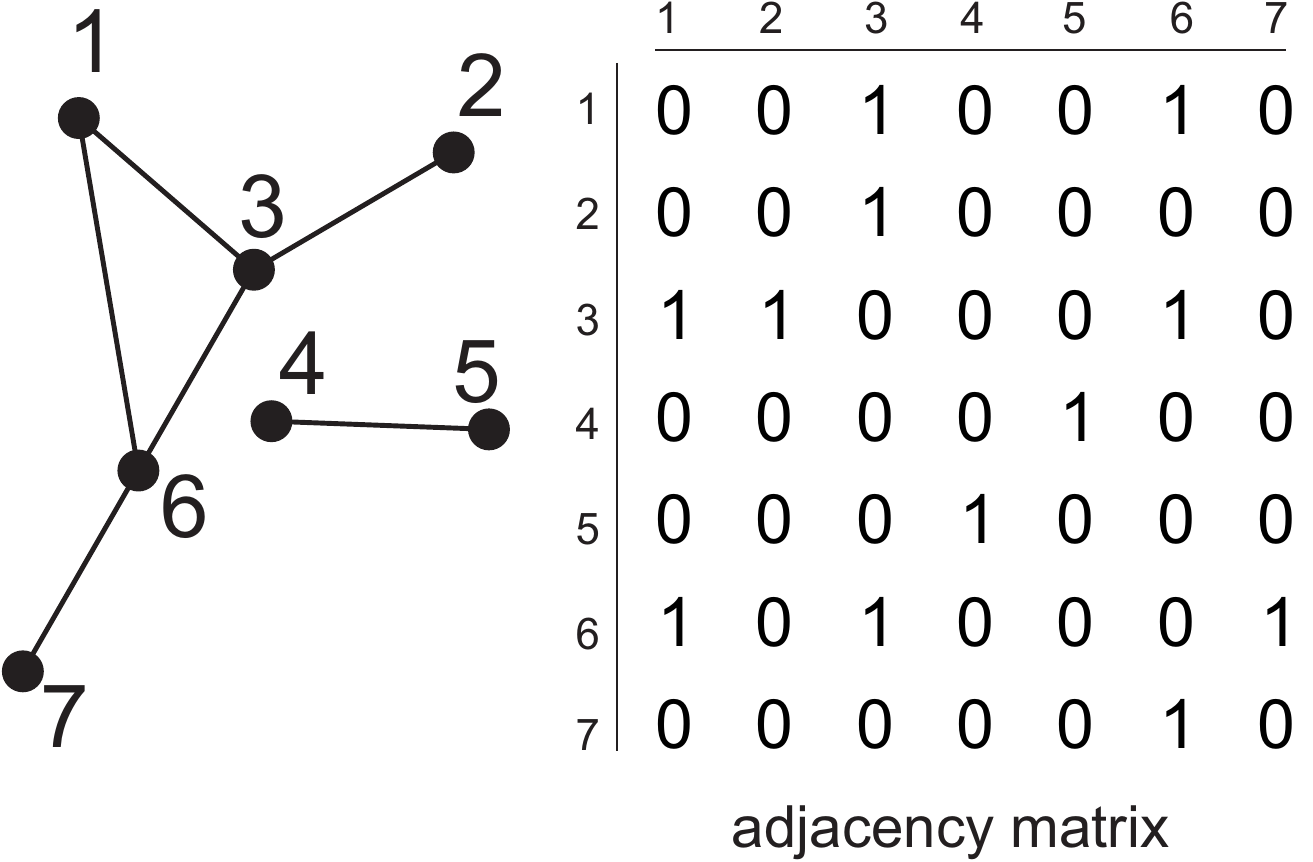} \hspace{1cm}
\includegraphics[width=7cm]{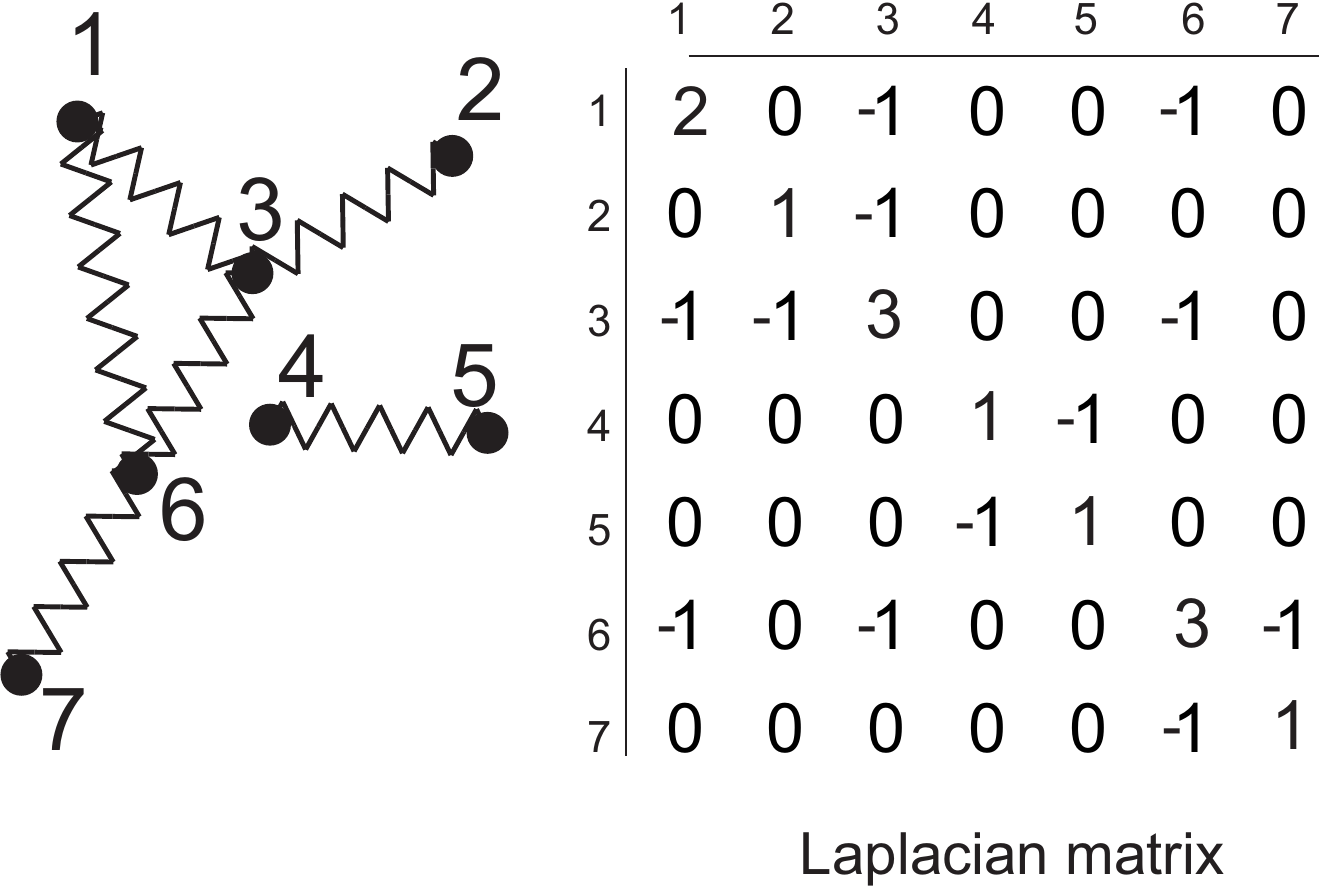}}
\caption{(a) Sample of a topological graph (collection of subgraphs) and its adjacency matrix;
(b) Elastic network corresponding to (a) and its Laplacian matrix.}
\label{fig:01}
\end{figure}

Consider the polymer network as a graph or a collection of graphs, see \fig{fig:01}a. Enumerate the
monomers of the $N$-atomic macromolecule by the index $i=1 \ldots N$. The adjacency matrix $A =
\{a_{ij}\}$ describing the topology (connectivity) of a polymer molecule is symmetric
($a_{ij}=a_{ji}$). Its matrix elements $a_{ij}$ take binary values, 0 and 1, such that the diagonal
elements vanish, i.e. $a_{ii}=0$, and for off-diagonal elements, $i\ne j$, one has $a_{ij}=1$, if
the monomers $i$ and $j$ are connected, and $a_{ij}=0$ otherwise.  The Laplacian matrix
$L=\{b_{ij}\}$ is by definition as follows: $b_{ij} = -a_{ij}$ for $i\ne j$, and $b_{ii} =
\sum_{j=1}^{N}a_{ij}$, as shown in \fig{fig:01}b, i.e. $L=d I-A$, where $d$ is the vector of vertex degrees of the graph and $I$ is the identity matrix. The eigenvalues $\lambda_n$ ($n=1,...,N$) of the symmetric matrix $L$ are real. For regular graphs (i.e for graphs with constant vertex degrees) the spectra of the adjacency and Laplacian matrices are uniquely connected to each other.

In physical literature the spectrum of adjacency matrix is interpreted as the set of resonance
frequencies, while the Laplacian spectrum defines the relaxation of the system. Thus, measuring the
response of the diluted solution of noninteracting polymer molecules on external excitations with
continuously changing wavelength, we expect to see the signature of eigenmodes in the spectral
density as peaks at some specific frequencies. Some other applications of adjacency and Laplacian
matrices in graph theory and optimization are thoroughly described in \cite{mohar1997some} and
\cite{chung1997spectral}.

Previously, in \cite{avetisov2015native} the eigenvalue density in ensembles of large sparse random
adjacency matrices was investigated. It has been demonstrated that the fraction of linear subgraphs at the percolation threshold is about 95\% of all finite subgraphs, and the distribution of linear
subgraphs (chains) is purely exponential. Analyzing in detail the spectral density of linear chain
ensembles, the authors claim in \cite{avetisov2015native} its \emph{ultrametric} nature and showed that near the edge of the spectrum, $\lambda_{max}$, the tails of the spectral density, $\rho(\lambda)$ exhibit a so-called ``Lifshitz singularity'', $\rho(\lambda)\sim e^{-c/\sqrt{\lambda_{max} -\lambda}}$, typical for the one-dimensional Anderson localization cite{lifshitz}. In \cite{avetisov2015native} the attention was paid to a connection of the spectral density to the modular geometry and, in particular, to the Dedekind $\eta$-function. In this work it was conjectured that ultrametricity emerges in any rare-event statistics and is inherit to generic complex sparse systems.

The rare-event statistics has many manifestations in physics \cite{planat} and biophysics
\cite{drosoph}. For example, the contact maps of individual DNA molecules in cell nuclei are sparse. Thus, experimenting with physical properties of highly diluted solutions of biologically active substances, one should pay attention to a very peculiar structure of background noise originating from the rare-event statistics of dissolved clusters. In this case, the peculiar shape of noise spectrum can be misinterpreted or at least can make the data incomprehensible
\cite{mairal2014sparse, peleg2010exploiting, vanden2012rare}. In order to make conclusions about any biological activity of regarded chemical substance, the signal from background noise should be clearly identified. From this point of view, the work \cite{rabadan} seems very interesting, since it represents an exceptional example of careful attention to ultrametric-like distributions in biological and clinical data.

In the current work we investigate in detail the spectral statistics (eigenvalue density of the
adjacency matrices) of branched polymer ensembles: (i) complete three-branching trees, and (ii)
$p$-branching star-like graphs. These structures represent special cases of branching polymers, known as \emph{dendrimers} \cite{furstenberg2012analytical}. Recall that dendrimers are regularly branching macromolecules with a tree-like topology. Computing the eigenvalues (resonances) of the adjacency matrix, and the degeneracy of these resonances in the polymer ensemble, we provide the information about the excitation statistics, analyze the tails of corresponding distributions and discuss the origin of clearly seen \emph{ultrametricity} \cite{mez,fra}. We obtain the results for adjacency matrices (but not for the Laplacian ones) to make our conclusions comparable with former research dealt with linear chain ensembles. Special attention we would like to pay to the works \cite{markelov1, markelov2}, where the relaxational spectra of dendrimer molecules have been analytically investigated and physical properties of corresponding dendrimers have been studied. Compared to the works \cite{markelov1, markelov2}, the goal of our paper is in some sense, contraversal: we would like to emphasize that dendrimers, as well as ensembles of linear polymers, and ensembles of sparse randomly branching clusters, share the generic hierarchical spectral statistics. From this point of view, the spectral statistics seen for dendrimers is \emph{not} uniquely determined by the topology of the graphs under consideration.

To make the content of the paper as self-consistent as possible, we remind briefly the notion of
ultrametricity. A metric space is a set of elements equipped by pairwise distances, $d(x,y)$
between elements $x$ and $y$. The metric $d(x,y)$ meets three requirements: i) non-negativity,
$d(x,y)>0$ for $x\neq y$, and $d(x,y)=0$ for $x=y$, ii) symmetry, $d(x,y)=d(y,x)$, and iii) the
triangle inequality, $d(x,z)\le d(x,y) + d(y,z)$. The concept of ultrametricity is related to a
special class of metrics, obeying the \emph{strong triangular inequality}, $d(x,z) \le
\max\{d(x,y), d(y,z)\}$.

The description of ultrametric systems in physical terms deals with the concept of hierarchical
organization of energy landscapes \cite{mez,fra}. A complex system is assumed to have a large
number of metastable states corresponding to local minima in the potential energy landscape. With
respect to the transition rates, the minima are suggested to be clustered in hierarchically nested
basins, i.e. larger basins consist of smaller basins, each of those consists of even smaller ones,
\emph{etc}. The basins of local energy minima are separated by a hierarchically arranged set of
barriers: large basins are separated by high barriers, and smaller basins within each larger one
are separated by lower barriers.

Since the transitions between the basins are determined by the passages over the highest barriers
separating them, the transitions between any two local minima obey the strong triangle inequality.
The ultrametric organization of spectral densities obtained in our work should be understood
exactly in that sense, if we identify the degeneracy of the eigenvalue with the height of the
barrier separating some points on the spectral axis. Ultrametric geometry fixes taxonomic (i.e.
hierarchical) tree-like relationships between elements and, speaking figuratively, is closer to
Lobachevsky geometry, rather than to the Euclidean one.

The selection of binary trees and stars for our study is not occasional. We are interested in the
question whether the hierarchical structure of spectral density emerges in polymer ensembles beyond
the linear statistics. Moreover, ensembles of full binary trees and stars allow for complete
analytic treatment, which makes them the first candidates beyond the rare-event statistics emerging
in linear ensembles. Varying the branching number of star-like graphs, we can interpolate between
linear chains and branching structures, attempting to understand the influence of non-linear
topology on spectral statistics.

It is essential to emphasize that we investigate ensembles of non-interacting polymer molecules.
This is ensured by the low density of polymers in the solution \cite{gennes}. Understanding how the
inter-molecular interactions change the spectral density of the polymer solution is a challenging
problem which yet is beyond the scopes of our investigations, however definitely will be tackled
later. Additional remarks should be made regarding the polymer size distribution in ensembles. It
is crucial that we study \emph{polydisperse} ensembles (i.e. ensemble of polymers of various
sizes). Here we suppose the probability $\PP(N)$ to find an $N$-atomic chain, to be exponential,
$\PP(N)~\sim~e^{-\mu N}$ ($\mu>0$ is some constant rate of joining monomers together in a course of
a polymer assembling). The choice of the exponential distribution is mainly motivated by the work
\cite{avetisov2015native} where such a distribution appeared naturally at the percolation threshold an follows from the random Bernoulli-type construction of long linear sequences. However, we can
consider any other distribution and the selection of the exponential one is mainly the question of
convenience: the comparison of new and old results in this case is much more straightforward.

\section{Graphs under consideration: full binary trees and stars}

In \cite{avetisov2015native} the authors discussed some statistical properties of polydisperse
ensembles of linear macromolecules and paid attention to two specific properties: i) the
singularity of the enveloping curve of spectral density at the edge of the spectrum, known as the
``Lifshitz tail'', and ii) the hierarchical organization of resonances in the bulk of the spectrum.
It was shown in \cite{avetisov2015native} that these properties are inherent to generic sparse
ensembles and can be viewed as number-theoretic manifestations of the rare-event statistics.
Whether they survive for ensembles of trees or stars is the question analyzed in present work.

\begin{figure}[ht]
\centerline{\includegraphics[width=12cm]{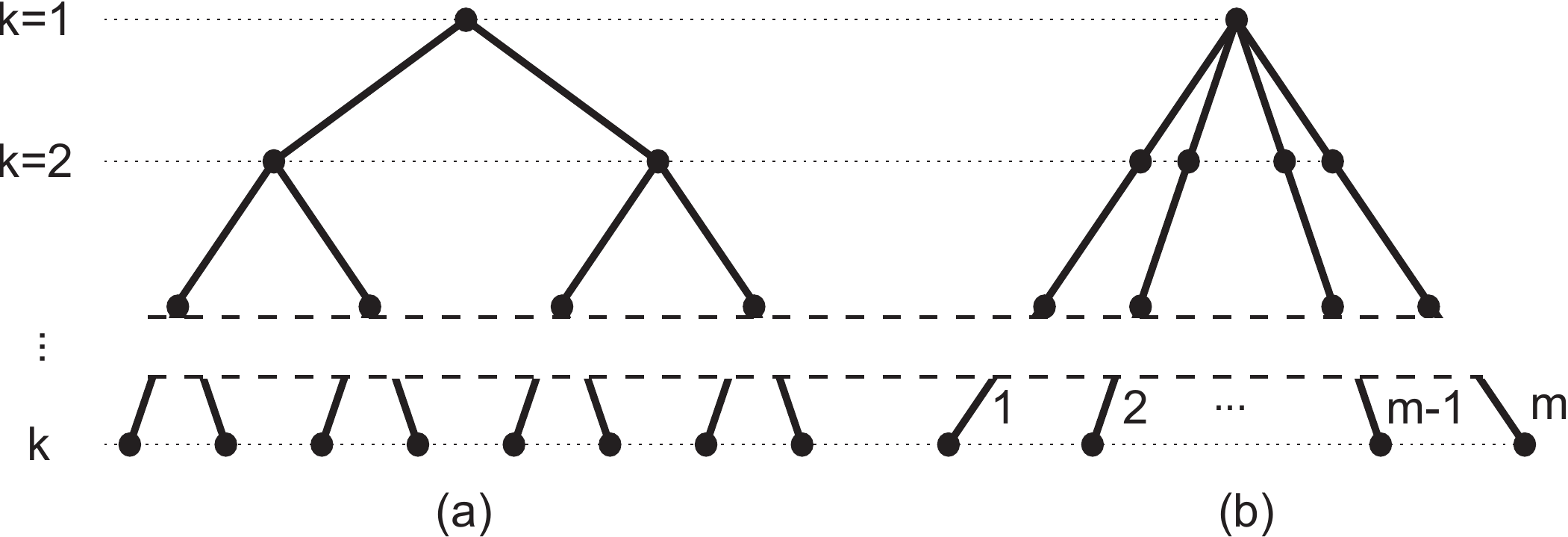}}
\caption{(a) Full binary tree; (b) a star-like graph with $p$ branches of $k$ nodes.}
\label{fig:trees1}
\end{figure}

Below we compute eigenvalues with corresponding multiplicities (degeneracies) of adjacency matrices
of full binary trees and star-like graphs (dendrimers), schematically depicted in
Figs.~\ref{fig:trees1}a,b. Then we perform averaging over ensembles of trees of particular topology
and determine the spectral density.

Let $\calT_k$ be a full binary tree of level $k$ shown in the Fig. \ref{fig:trees1}a. The shortest
path from the root of $\calT_k$ to any leaf is $k - 1$, and the tree has $2^k-1$ vertices in total.
Let $\bT_k$ be the adjacency matrix of $\calT_k$. The spectrum of an individual tree $\bT_k$ has
exactly ${\cal N}^{tree}_k=2^k-1$ eigenvalues, all real. Let $\rho_k (\lambda): \RR \rightarrow \RR$ be the spectral
density of $\bT_k$,
\be
\rho_k (\lambda) = \frac{\tM_k(\lambda)}{{\cal N}^{tree}_k},
\label{eq:01}
\ee
where $\lambda$ is an eigenvalue, and $\tM_k(\lambda)$ is its multiplicity. We investigate the ensemble of such trees with the exponential distribution,
\be
\PP_{\mu}(k) = C e^{-\mu k}, \qquad \sum_{k=1}^{\infty} \PP_{\mu}(k)=1,
\label{eq:02}
\ee
where $C=e^{\mu} - 1$, and $\mu$ is a parameter. The spectral density, $\rho(\lambda)$ of the
ensemble is the quotient of multiplicity expectation and tree size expectation
\be
\rho(\lambda) = \frac{\sum\limits_{k = 1}^{\infty} \PP_{\mu}(k)\ \tM_k(\lambda)}{\sum\limits_{k =
1}^{\infty} \PP_{\mu}(k)\ {\cal N}_k^{tree}}.
\label{eq:03}
\ee

For a star graph depicted in the Fig. \ref{fig:trees1}b we state the problem similarly. Let
$\calS_{k,p}$ be a star graph constructed by gluing $p$ linear graphs (chains) at one point, all
extended up to the level $k$. The graph has $p(k-1) + 1$ vertices in total, and its adjacency
matrix is $\bS_{k,p}$ with the total number of eigenvalues ${\cal N}_{k,p}^{star} =p(k-1) + 1$. Let
$g_{k,p} (\lambda): \RR \rightarrow \RR$ be the spectral density of $\bS_{k,p}$,
\be
g_{k,p} (\lambda) = \frac{\tM_{k,p}(\lambda)}{{\cal N}_{k,p}^{star}},
\label{eq:04}
\ee
where $\lambda$ is an eigenvalue and $\tM_{k,p}(\lambda)$ is its multiplicity. The probability of a star to have parameters $k$ and $p$ is $\PP(k|p) \PP (p)$. Specifically, we consider the ensemble of stars with fixed $p$, where $k$ is distributed as in
(\ref{eq:02}):
\be
\PP_{\mu}(k|p) = C e^{-\mu k}, \qquad \sum_{k=1}^{\infty} \PP_{\mu}(k|p) = 1, \qquad \sum_{p\ge 1}
\PP_{\mu}(p) = 1.
\label{eq:05}
\ee
Thus, the spectral density $g_p(\lambda)$ of the $p$-branch star ensemble is
\be
g_p(\lambda)= \frac{\sum\limits_{k=1}^{\infty} \PP_{\mu}(k|p) \tM_{k, p} (\lambda)}{\sum\limits_{k
= 1}^{\infty} \PP_{\mu}(k|p)\ {\cal N}_{k,p}^{star}}.
\label{eq:06}
\ee
We suppose that $p\ge 3$, since $p = 1, 2$ corresponds to linear chains.

\section{Results}

Below we provide numeric and analytic results for spectral densities of dendrimer ensembles. Numeric simulations allow us to get a visual representation of the investigated objects in order to make plausible conjectures about individual graph spectra as well as to understand generic features of the corresponding spectral densities of the systems under consideration.

\begin{figure}[!t]
\centering \subfloat[Spectral density of the full binary tree
$\calT_{10}$.]{\includegraphics[width=0.33\textwidth]{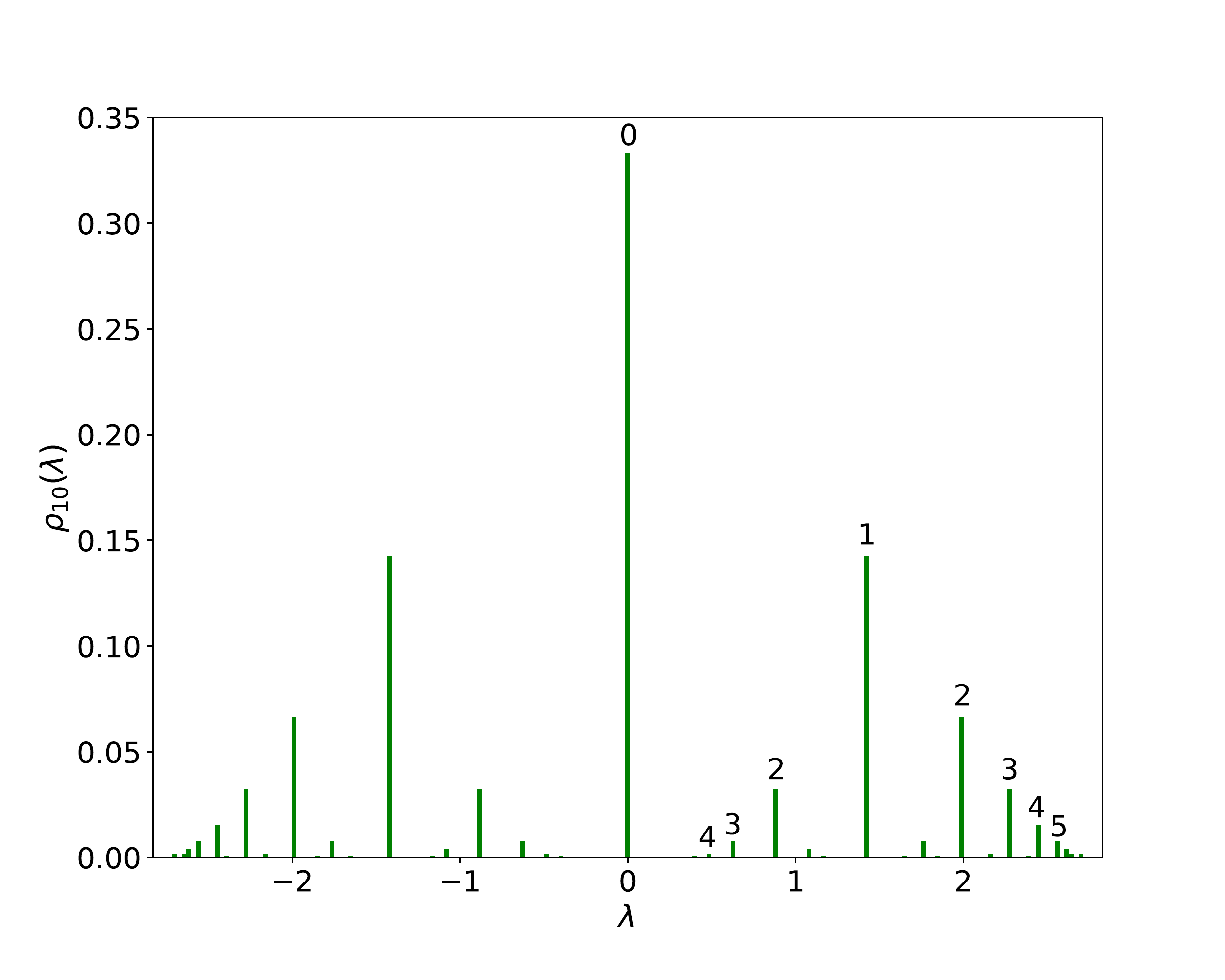}
\label{fig:fb_10}}
\centering \subfloat[Spectral density of the full binary tree $\calT_{10}$ in the $\log-\log$ coordinates (to be compared to the Fig.2b from \cite{markelov2}).] {\includegraphics[width=0.33\textwidth]{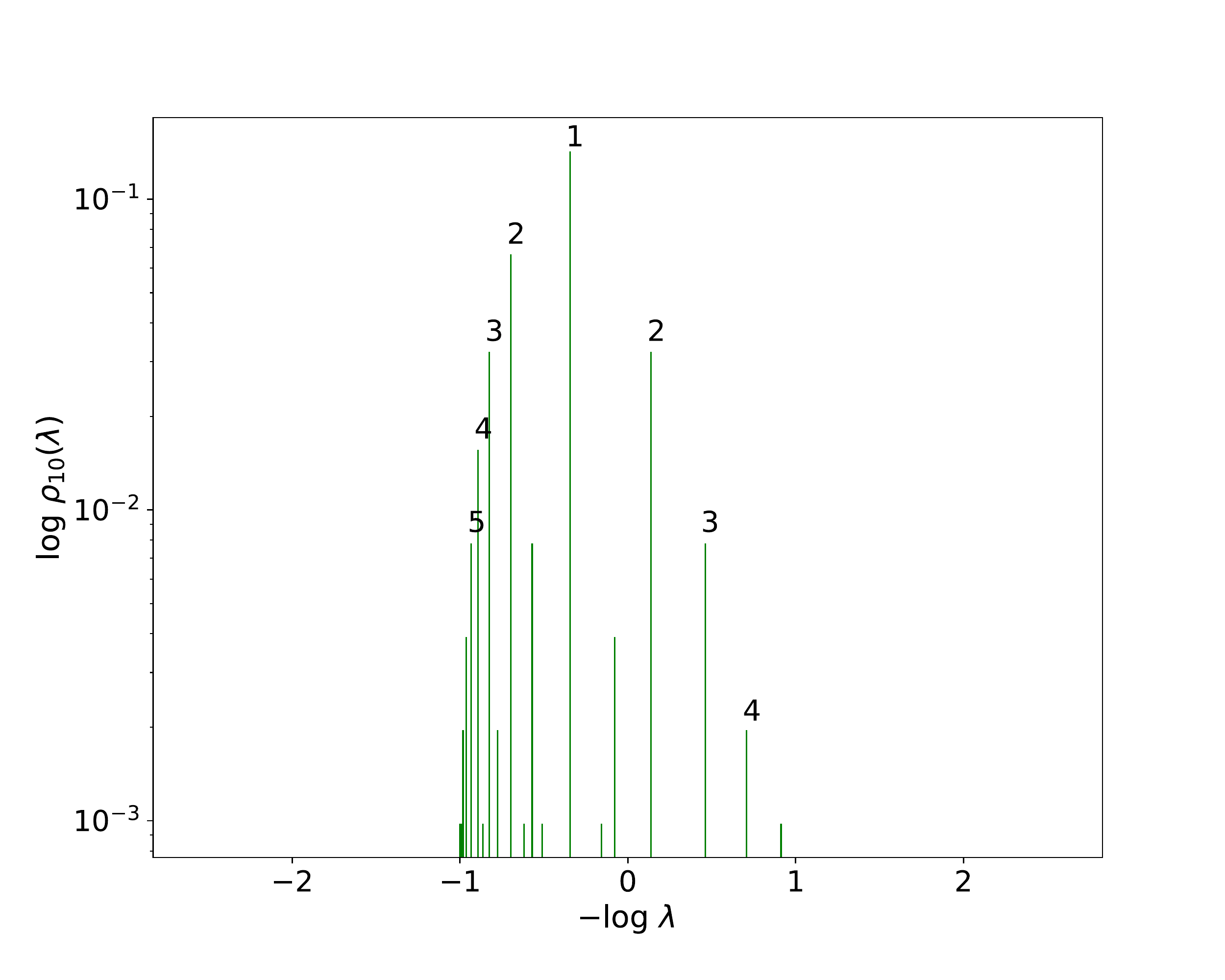}
\label{fig:fb_10b}}
\subfloat[Spectral density of the exponential ensemble with parameter $\mu = 0.5$]{\includegraphics[width=0.33\textwidth]{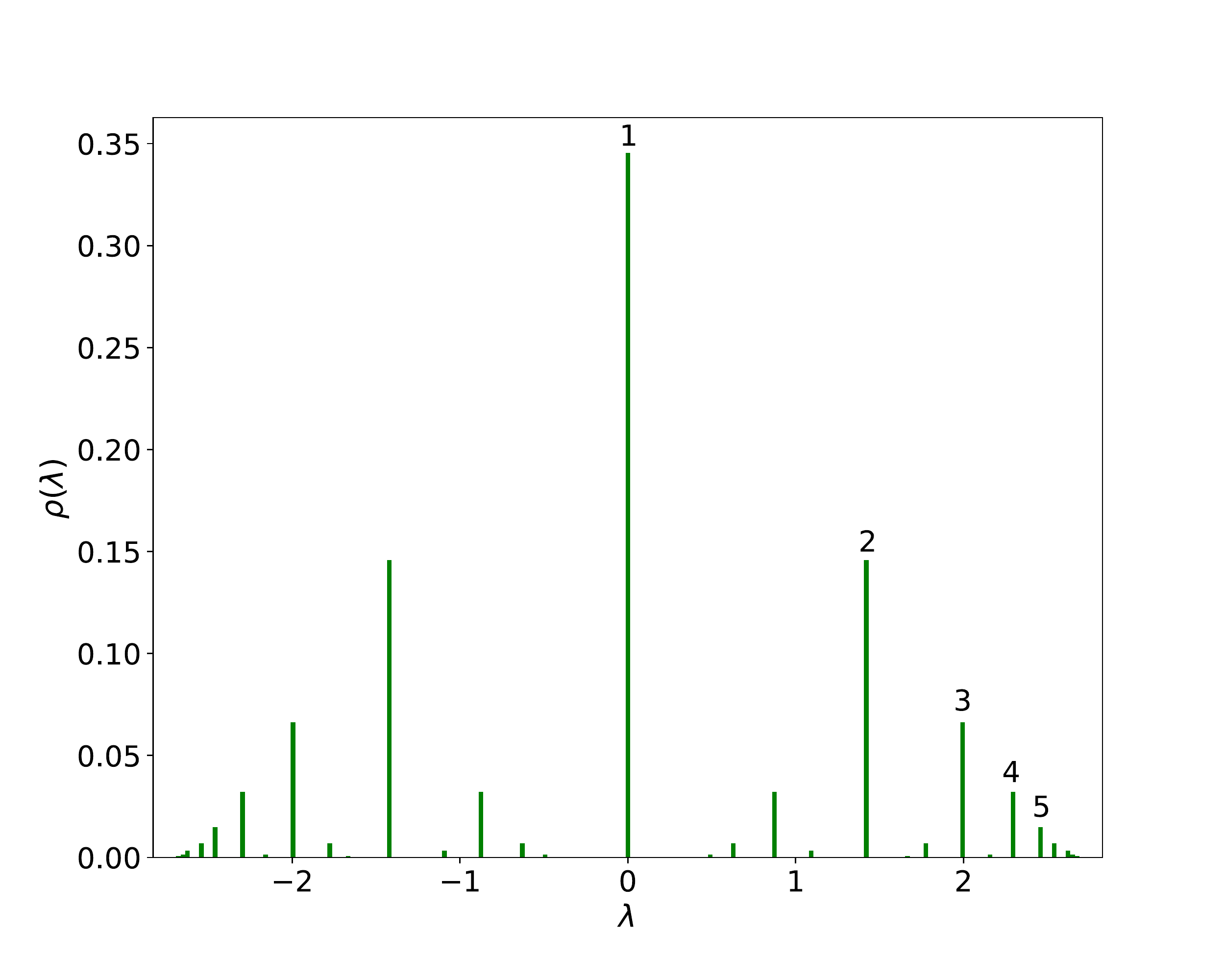}
\label{fig:fb_ens}}
\caption{Eigenvalue distribution of full binary trees. Axis X is eigenvalue, axis Y is its frequency.}
\label{fig:fb_all}
\end{figure}

The main steps of the numeric algorithm are as follows. We calculate eigenvalues of $\bT_k$ or $\bS_{k, p}$ with corresponding multiplicities. The computational complexity of eigenvalue calculation is $O(n^3)$, where $n$ the number of vertices. Note that accumulation of small computational errors can lead to inaccurate results. That is why we construct a histogram instead of treating each eigenvalue separately. Specifically, we divide the axis of eigenvalues into the intervals of length $\Delta$ and construct the piecewise constant function $\hat{f}_k (\lambda): \RR \rightarrow \RR$. On a particular segment the function $\hat{f}_k (\lambda)$ is equal to the number of eigenvalues in this segment:
\be
\hat{f}_k(\lambda) = \frac{\#\left\{\lambda_i \in \left[\lambda_{min} + l\Delta; \lambda_{min} +
(l+1)\Delta\right]\right\}}{{\cal N}},\quad \forall \lambda \in \left[\lambda_{min} + l\Delta;
\lambda_{min} + (l+1)\Delta\right],
\label{eq:07}
\ee
where $\lambda_i$ is an eigenvalue, and $l$ is the index of the segment. Since the maximal eigenvalue is bounded for both types of trees (see \cite{grosberg2015statistics})
\begin{equation}
|\lambda^{\tree}_{\max}| \le 2\sqrt{p - 1},
\label{eq:upper_bound}
\end{equation}
we only scan the support $\Big[-|\lambda^{\tree}_{\max}|; |\lambda^{\tree}_{\max}|\Big]$. Finding the value of $\Delta$ that provides for the most representative histogram is a separate technical question which is not addressed here. With such histograms (for both full binary trees and stars) for various levels one can easily reconstruct the desired spectral density by performing convolutions of the functions $\rho_k(\lambda)$ and $g_{k,p}(\lambda)$ with the distribution functions as prescribed by \eq{eq:03} and \eq{eq:06}. The results of our computations of spectral densities (i.e. the densities of eigenvalues) for individual tree and for an ensemble of exponentially weighted trees are summarized in \fig{fig:fb_all}.

We would like to pay attention to the \fig{fig:fb_10b} which represents the same histogram as in \fig{fig:fb_10} replotted in the $\log-\log$ coordinates. This figure should be compared to Fig.2b of the work \cite{markelov2} where the relaxational spectrum of a specific dendrimer has been computed. The similarity between the spectral densities of: i) individual dendrimers, ii) ensembles of dendrimers, and iii) linear chains with exponential distribution in lengths, allows us to conclude that the spectral density does not reflect the topology of the system under consideration, demonstrating rather generic features of one-dimensional systems with quenched disorder.

\subsection{Full binary trees: numerics}

The number of vertices of a full binary tree grows exponentially with the level, while the structure of the spectrum is very stable, which is not surprising since the full binary tree is a self-similar structure. Note that the maximal vertex degree of a full binary tree is $p = 3$, so, according to \eq{eq:upper_bound}, $|\lambda^{\tree}_{\max}| \le 2\sqrt{2}$ for any $k$.

The \fig{fig:fb_all} shows typical spectral densities, namely: the \fig{fig:fb_10} depicts the spectral density of a \emph{single} full binary tree with $k=10$ levels, and the \fig{fig:fb_ens} demonstrates the spectral density of the full binary tree ensemble (averaged over $1000$ realizations). In simulations we have generated $k$ with the exponential distribution \eq{eq:02}, calculated eigenvalues of the corresponding particular tree and added all values from different trees to the list, thus considering a forest of full binary trees with prescribed level distribution. Note the similarity between \fig{fig:fb_10} and \fig{fig:fb_ens}.

For better clarity we provide numeric data for main peak multiplicities ($k = 8, 9, 10$) in the Table \ref{tab:1}. The peaks are enumerated as shown in \fig{fig:fb_all}. Note that the diagonals in this table are relatively stable. In the Table \ref{tab:2} we present the main peak frequencies for $k = 8, 9, 10$ and also for exponentially distributed ensembles with $\mu = 0.5$ and $\mu = 0.8$. Note that frequencies in lines with fixed peak number are also relatively stable regardless of tree size or even whether it is an individual tree or an ensemble.

\begin{table}[ht]
\subfloat[]{ \label{tab:1}
\begin{tabular}{|c||c|c|c|}
\hline
Peak No. & $k = 8$ & $k = 9$ & $k = 10$ \\
\hline \hline
$0$ & $85$ & $171$ & $341$ \\
\hline
$1$ & $37$ & $73$ & $146$ \\
\hline
$2$ & $17$ & $34$ & $68$ \\
\hline
$3$ & $8$ & $17$ & $33$ \\
\hline
$4$ & $4$ & $8$ & $16$ \\
\hline
\end{tabular}
} \hspace{0.5cm} \subfloat[]{ \label{tab:2}
\begin{tabular}{|c||c|c||c|c|c|}
\hline
Peak No. & $\mu=0.5$ & $\mu=0.8$ & $k = 8$ & $k = 9$ & $k = 10$\\
\hline \hline
$0$ & $0.346$ &$0.392$& $0.333$ & $0.335$ & $0.333$\\
\hline
$1$ &$0.146$ &$0.152$& $0.145$ & $0.143$ & $0.143$\\
\hline
$2$ & $0.067$ &$0.062$& $0.067$ & $0.067$ & $0.068$\\
\hline
$3$ & $0.032$ &$0.029$& $0.031$ & $0.033$ & $0.032$\\
\hline
$4$ & $0.015$ &$0.012$& $0.016$ & $0.016$ & $0.016$\\
\hline
\end{tabular}
}
\caption{(a) Individual trees: peak multiplicities for various $k$ in the enveloping series; (b)
Ensemble of trees with $\mu = 0.5$ and $\mu = 0.8$ and individual trees ($k = 8,9,10$): peak frequencies in the enveloping series.}
\end{table}

\subsection{Full binary trees: theory}

\subsubsection{Spectral density of a single full binary tree}

Let $\calT_k$ is a full binary tree and $\bT_k$ is its adjacency matrix. If the vertices are enumerated linearly through levels $1$ to $k$, $\bT_k$ takes the following form:
\be
\bT_k = \begin{bmatrix}
0\ & 1\ & 1 \ & 0 \ & 0 \ & 0 \ & 0 \ \\
1 & 0 & 0 & 1 & 1 & 0  & 0 \\
1 & 0 & 0 & 0 & 0 & 1 & 1 \\
0 & 1 & 0 & 0 & 0  & 0 & 0 \\
0 & 1& 0 & 0 & 0 & 0 & 0 \\
0 & 0 & 1 & 0 & 0 & 0 & 0 \\
0 & 0 & 1 & 0 & 0 & 0 & 0 \\
\end{bmatrix}
\ee
Here $k = 3$, and for other $k$ the structure is the same. However, we will not operate with this form as it does not provide for clear analysis. Instead, we introduce alternative notation.

\begin{figure}[ht]
\centerline{\includegraphics[width=7cm]{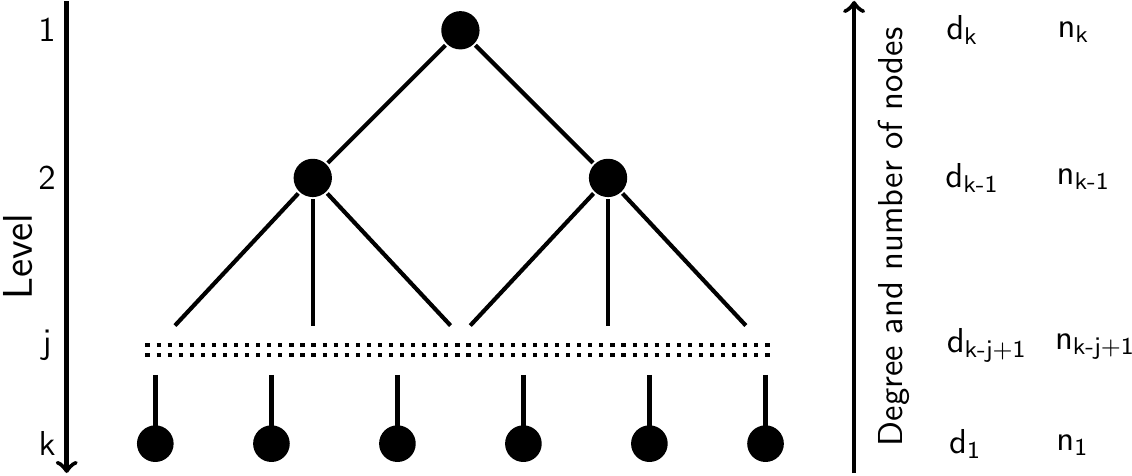}}
\caption{A generalized Bethe tree.}
\label{fig:trees2}
\end{figure}

Following \cite{rojo2005spectra,rojo2007explicit} denote $\calB_k$ a generalized Bethe tree of $k$ levels, shown in~\fig{fig:trees2}.

\newtheorem{mydef}{Definition}
\begin{mydef}
A generalized Bethe tree $\calB_k$ is a rooted non-weighted and non-directed tree with vertex degree depending on the distance from the root only. All vertices in the same level have equal degree. The number of levels equals $k$.
\end{mydef}

Enumerate the levels down-up by index $j=1,...,k$. The root is located at $j=1$ and the number of vertices at the level $j$ is $n_{k-j+1}$, all of them of equal degree $d_{k-j+1}$. In particular, $d_k$ is the degree of the root, $n_k = 1$ and $n_1$ is the number of terminal vertices (``leaves'').

Let $\bA(\calB_k)$ be the adjacency matrix of $\calB_k$ and $\sigma(\bA(\calB_k))$ -- its spectrum. Define also $\bd = (1, d_2, \ldots, d_k)$ and $\Omega = \{j:\ 1 \le j \le k-1,\ n_j > n_{j+1}\}$. The following theorem (the Theorem 1 of \cite{rojo2007explicit}) is the key ingredient of our study:

\newtheorem{th1}[counter]{Theorem}
\begin{th1} If $\bA_j(\bd)$ is the $j \times j$ leading
principal submatrix of the $k \times k$ symmetric tridiagonal matrix
\be
\bA_{k}(\bd)= \scalemath{0.90}{
\begin{bmatrix}
0 & \sqrt{d_2-1} && \\ \sqrt{d_2-1} & 0 & \sqrt{d_3-1} && \\ & \sqrt{d_3-1} & 0 &&  \\
\ldots & \ldots & \ldots & \ldots & \ldots & \ldots \\ &&& 0 & \sqrt{d_k-1} &  \\
 &&& \sqrt{d_k-1} & 0 & \sqrt{d_k} \\  & & & & \sqrt{d_k} & 0
\end{bmatrix}}
\label{eq:08}
\ee
then
\begin{enumerate}
\item $\sigma(\bA(\mathcal{B}_k))= \left(\bigcup_{j \in \Omega} \sigma(\bA_j(\bd))\right)
\cup \sigma(\bA_k(\bd))$.
\item The multiplicity of each eigenvalue of the matrix $\bA_j(\bd)$ as an eigenvalue of
$\bA(\mathcal{B}_k)$ is $(n_j-n_{j+1})$ for $j \in \Omega$, and eigenvalues of $\bA_k(\bd)$ as
eigenvalues of $\bA(\mathcal{B}_k)$ are simple.
\end{enumerate}
\end{th1}

This theorem allows us to investigate a simple tridiagonal matrix instead of the full adjacency matrix of a tree. The  size of corresponding tridiagonal matrix equals to the number of levels, which is exponentially smaller than the number of vertices. Now we formulate and prove the theorem about the spectrum of trees under consideration.

\newtheorem{th2}[counter]{Theorem}
\begin{th2}
Let $\calT_k$ be the full binary tree of $k$ levels, $\sigma(\calT_k)$ - its spectrum, $\tM_k(\lambda)$ and $\rho_k(\lambda)$ are the multiplicity and the frequency of the eigenvalue $\lambda \in \sigma(\calT_k)$, then:
\begin{enumerate}
\item $\sigma(\calT_k) = \bigcup_{j = 1}^k \bigcup_{i = 1}^{j}\left\lbrace 2\sqrt{2} \cos \frac{\pi i}{j + 1}\right\rbrace.$
\item $\tM_k(\lambda) = \frac{2^k - 2^{k\ \mod (j+1)}}{2^{j+1} - 1} + \II\{k \equiv j\ \mod (j+1)\}.$
\item $\rho_k(\lambda) = \frac{1}{2^k-1}
\left(\frac{2^k - 2^{k\ \mod (j+1)}}{2^{j+1} - 1} + \II\{k \equiv j\ \mod (j+1)\}\right).$
\end{enumerate}
\end{th2}

\begin{proof}

The graph $\calT_k$ has $n_{k-j+1}=2^{j - 1}$ vertices at the level $j$, all of degree $d_j = 3$ except for the level $1$ ($d_1 = 1$) and the root ($d_k = 2$). According to \yury{t}heorem 1, its spectrum is the
union of submatrix spectra:
\be
\sigma (\bT_k) = \bigcup_{j = 1}^k \sigma (\bA_j),
\label{eq:09}
\ee
where $\bA_j$ is the following:
\be
\bA_j = \sqrt{2}\begin{bmatrix}
0 & 1 & & \\ 1 & 0 & 1 & & \\
\cdots & \cdots & \cdots & \cdots & \cdots\\
 & & 1 & 0 & 1 \\ & & & 1 & 0 \\
\end{bmatrix}
\label{eq:10}
\ee

Let $\hat{\bA}_j = \dfrac{1}{\sqrt{2}}\ \bA_j$ and $\hat{\lambda} = \dfrac{1}{\sqrt2}\ \lambda$. Note that $\hat{\bA}_j$ is actually the adjacency matrix of the linear chain.
The characteristic polynomial $F_j~=~\det (\hat{\bA}_j~+~\hat{\lambda} \bE)$ of the matrix
$\hat{\bA}_j$ satisfies the following recurrence equation:
\be
\left\{\begin{array}{c}
F_{j+1} = \hat{\lambda} F_j - F_{j-1}, \medskip \\
F_0 = 1, \; F_1 = \hat{\lambda}.
\end{array} \right.
\label{eq:11}
\ee
and the corresponding characteristic equation $\mu^2 - \mu\hat{\lambda} + 1 = 0$ has roots
\be
\mu_\pm = \frac{\hat{\lambda} \pm \sqrt{\hat{\lambda}^2 - 4}}{2}.
\label{eq:12}
\ee
This allows us to reconstruct the explicit form of $F_j$:
\be
F_j = \frac{1}{\sqrt{\hat{\lambda}^2 - 4}}\ \mu_+^{j+1} - \frac{1}{\sqrt{\hat{\lambda}^2 - 4}}\
\mu_-^{j+1}.
\label{eq:13}
\ee
Assuming $\hat{\lambda} \in (-2;2)$, we solve $F_j = 0$. Its roots are:
\be
\hat{\lambda} = 2\cos \frac{\pi i}{j + 1}, \quad i = 1 \ldots j.
\label{eq:14}
\ee
There are $j$ of them, and as $F_j$ is a polynomial of degree $j$, there are no other roots.
Thus, the eigenvalues of the matrix $\bA_j$ are
\be
\lambda = 2 \sqrt{2} \cos \frac{\pi i}{j + 1},\quad i = 1 \ldots j,
\label{eq:15}
\ee

Collecting all intermediate results, we arrive at the following explicit expression for the spectrum of an individual tree $\calT_k$:
\be
\sigma(\bT_k) = \bigcup_{j = 1}^k \bigcup_{i = 1}^{j}\left\lbrace 2\sqrt{2} \cos \frac{\pi i}{j +
1}\right\rbrace.
\label{eq:spec_Tk}
\ee
Note that the spectrum of an individual tree is the same as the scaled spectrum of a set of linear chains, and level defines here the maximal length of a linear chain.

Next, compute the multiplicities. According to the second statement of \yury{t}heorem 1, the contribution of the $j^{th}$ principal submatrix is:
\be
m_j = n_j - n_{j+1} = 2^{k - j - 1} \quad (j = 1 \ldots k - 1), \quad m_k = 1.
\label{eq:17}
\ee
Recall that the multiplicity of the eigenvalue $\lambda$ is $\tM_k(\lambda)$. Then we have:
\be
\tM_k(\lambda) = \sum_{j = 1}^k m_j\, \II\{\lambda \in \sigma(\bA_j)\},
\label{eq:defMk}
\ee
where $\II\{\cdot\}$ is the indicator function:
\[
\II\{\cdot\} = \left\{\begin{array}{ll} 1, & \mbox{if the condition is true} \medskip \\
0, & \mbox{otherwise} \end{array} \right.
\]

Eigenvalue $\lambda$ belongs to $\sigma(\bA_j)$ if there exists an integer $i$ such that
\be
\lambda = 2\sqrt{2}\ \cos \frac{\pi i}{j + 1}.
\label{eq:19}
\ee
Supposing that $\frac{i}{j+1}$ is irreducible, it contributes to the spectrum as an eigenvalue of
every $(j+1)^{\rm th}$ principal submatrix:
\be
2\sqrt{2}\cos \frac{\pi i}{j + 1},\; 2 \sqrt{2}\cos \frac{2\pi i}{2(j + 1)},\; \ldots, \; 2
\sqrt{2} \cos \frac{(n-1)\pi i}{(n-1)(j + 1)}, \; 2 \sqrt{2}\cos \frac{n\pi i}{n (j + 1)}
\label{eq:20}
\ee
where $n = \left\lfloor \frac{k+1}{j+1} \right\rfloor$. The respective multiplicities are:
\be
2^{k-(j+1)},\; 2^{k-2(j+1)}, \; \ldots,\; 2^{k-(n-1)(j+1)}, \; m_n
\label{eq:21}
\ee
where $m_n$ depends on $k$ and $j$. It is also important if $\lambda$ is an eigenvalue of the matrix itself as in this case the multiplicity is $1$, otherwise it is a difference of two powers of two.
\be
m_n =
\begin{cases}
1,& \text{if } k \equiv j\ \mod (j+1)\\
2^{k - n(j+1)} = 2^{k\ \mod (j+1)}, & \text{otherwise}
\end{cases}
\label{eq:22}
\ee
Summing the geometric series, we arrive at the following expression for the multiplicity of
$\lambda = 2\sqrt{2}\cos \dfrac{\pi i}{j + 1}$, defined in (\ref{eq:defMk}):
\be
\tM_k(\lambda) = \begin{cases}
\dfrac{2^k + 2^j - 1}{2^{j+1} - 1}, & \text{if } k \equiv j\ \mod (j+1) \medskip\\
\dfrac{2^k - 2^{k\ \mod (j+1)}}{2^{j+1} - 1}, & \text{otherwise}
\end{cases}
\label{eq:23}
\ee
or
\be
\tM_k(\lambda) =
\frac{2^k - 2^{k\ \mod (j+1)}}{2^{j+1} - 1} + \II\{k \equiv j\ \mod (j+1)\}.
\label{eq:24}
\ee

The spectral density, $\rho_k$, of an individual tree is
\be
\rho_k(\lambda) = \frac{\tM_k(\lambda)}{{\cal N}_k^{\tree}} = \frac{1}{2^k-1}
\left(\frac{2^k - 2^{k\ \mod (j+1)}}{2^{j+1} - 1} + \II\{k \equiv j\ \mod (j+1)\}\right).
\label{eq:rho_k}
\ee
That finishes the proof of this theorem.
\end{proof}

Now we know exactly all eigenvalues and their multiplicities or frequencies shown in \fig{fig:fb_all}. An important remark is that an individual full binary tree possesses the spectral density, which is identical to the spectral density of the exponentially weighted ensemble of linear chains up to the scaling factor (the locations of the peaks are multiplied by $\sqrt{2}$). We believe that this similarity should be taken into account in view of the works \cite{markelov1,markelov2} about relaxational properties of dendrimers: looking at the spectra, we cannot distinguish the full binary tree from the diluted polydisperse solution of trees (dendrimers), and even linear macromolecules, whose lengths are distributed exponentially.

Apart from that, we are interested in the asymptotic behavior of two peak series defining the envelope of the spectral density. Generically, the behavior of tails of the spectral density of disordered system provides the information about the extreme value statistics of top eigenvalues. For example, the Wigner semicircle for the eigenvalue density of an ensemble of Gaussian unitary matrices \cite{mehta} is violated at the edge of the spectrum and the statistics of the highest (top) eigenvalue shares the so-called Tracy-Widom distribution \cite{tracy}. So, it seems challenging to derive the behaviors of the spectrum density near the spectral edge and in the bulk and compare these situations to some known cases.

Define the ``outer curve"\ $\rho_k^{out}(\lambda)$ as the series of main peaks, sequentially enumerated as shown in \fig{fig:fb_all}. It is also the series of maximal eigenvalues of levels from $2$ to $k$:
\be
\lambda_j^{out} = 2 \sqrt{2}\ \cos \frac{\pi}{j + 1}, \quad j = 2, \ldots, k.
\ee
The second series defines the ``inner curve"\ $\rho_k^{in} (\lambda)$, which is the enveloping curve of the peaks descending from the second main peak to zero. In other words, it is the series of minimal positive eigenvalues:
\be
\lambda_i^{in} = 2 \sqrt{2}\cos\frac{\pi j}{2j + 1} = 2 \sqrt{2}\sin \frac{\pi}{2(2j + 1)}, \quad j
= 1, \ldots, \left\lfloor\frac{k}{2}\right\rfloor.
\ee

\newtheorem{th3}[counter]{Theorem}
\begin{th3} Let $\calT_k$ be the full binary tree of level $k$, then the inner curve $\rho_k^{in}(\lambda)$ and the outer curve $\rho_k^{out}(\lambda)$ of its spectral density $\rho_k(\lambda)$ have exponential asymptotics:
\begin{enumerate}
\item $\rho_k^{out} (\lambda) \sim \exp \left(-\frac{A}{\sqrt{\lambda_{max}-\lambda}}\right)$ at $\lambda \rightarrow \lambda_{max}$,
\item $\rho_k^{in}(\lambda) \sim \exp
\left(-\frac{B}{\lambda}\right)$ at $\lambda \rightarrow 0$,
\end{enumerate}
where $A$ and $B$ are some positive absolute constants.
\end{th3}

\begin{proof}

Considering the asymptotics of the tail of the ``outer curve"\ ,  or its behavior as $\lambda \rightarrow \lambda_{max} = 2\sqrt{2}$, is the same as considering $k\gg 1$ and $j \gg 1$.

For such values we use the Taylor expansion of the $\cos$-function
\be
\lambda = 2 \sqrt{2}\cos\frac{\pi}{j + 1} = 2\sqrt2\left(1 - \frac{\pi^2}{2 (j+1)^2}\right) +
o\left(\frac{1}{(j+1)^2}\right)
\label{eq:28}
\ee
Inverting \eq{eq:28}, we get
\be
j+1 = \frac{\sqrt[4]{2} \pi}{\sqrt{2\sqrt2 - \lambda}} + o\left(\frac{1}{(j+1)^2}\right)
\label{eq:29}
\ee
Both $i$ and $\lambda$ are in fact discrete and this expression only provides for a good approximation. Substituting \eq{eq:29} into \eq{eq:rho_k}, we obtain the asymptotic expression of the tail $\rho_k^{out}(\lambda)$ near the spectrum edge, i.e. at
$|\lambda|\to\lambda_{max}$:
\be
\rho_k^{out} (\lambda) = \exp \left(-\frac{ \sqrt[4]{2}\pi \ln 2}{\sqrt{2\sqrt2-|\lambda|}}\right)(1 + o(1))
\sim \exp \left(-\frac{A}{\sqrt{\lambda_{max}-|\lambda|}}\right), \quad A =  \sqrt[4]{2} \pi \ln 2.
\label{eq:30}
\ee


Now consider the asymptotics of the inner curve at $j\gg 1$, $k \gg 1$ in terms of indices, and $\lambda \rightarrow 0$ in terms of eigenvalues.
Apply the Taylor expansion of the $\sin$-function:
\be
\lambda = 2 \sqrt{2} \sin \frac{\pi}{2(2j + 1)} = \frac{\pi}{\sqrt2 (2j+1)} +
o\left(\frac{1}{2j+1}\right);
\label{eq:31}
\ee

Inverting this expression and substituting it into \eq{eq:rho_k} in the same manner we did for the outer curve, we arrive at the following
asymptotics of the inner curve (at $\lambda\to 0$):
\be
\rho_k^{in}(\lambda) \sim \exp \left(-\frac{\sqrt2 \pi \ln 2}{|\lambda|}\right) = \exp
\left(-\frac{B}{|\lambda|}\right), \quad B = \sqrt2 \pi \ln 2.
\label{eq:32}
\ee
\end{proof}


The behavior \eq{eq:30} signals the existence of so-called ``Lifshitz tail'' typical for the edge of the spectral density in physics of one-dimensional disordered systems \cite{lifshitz,pastur}. The emergence of similar behavior in deterministic tree-like system allows to suggest that regular tree-like graphs could be served as models which mimic some properties of ensembles of one-dimensional disordered systems like, for example, the one-dimensional disordered alloys (see \cite{alloys} for presenting the issue and a number of related references).


\subsubsection{Spectral density of full binary tree ensembles}

In this section we derive the eigenvalues and their frequencies for the ensemble of full binary trees, which is a set of full binary trees with sizes distributed exponentially. It turns out that, indeed, the results are pretty similar to the case of a single tree.

\newtheorem{th4}[counter]{Theorem}
\begin{th4} Let $\sigma^{\rm ensemble}$ be the spectrum of the full binary tree ensemble with level distributed exponentially, and $\rho(\lambda)$ be the spectral density of such ensemble, then

\begin{enumerate}
\item $\sigma^{\rm ensemble} = \bigcup_{j = 1}^{\infty} \bigcup_{i = 1}^{j}\left\lbrace 2\sqrt{2} \cos \frac{\pi i}{j + 1}\right\rbrace,$
\item For every $\lambda \in \sigma^{\rm ensemble}$, i.e. $\lambda = 2\sqrt2 \cos \frac{\pi i}{j+1}$, where $i$ and $j$ are coprime, $\rho(\lambda) = \dfrac{(e^{\mu} - 1)^2}{e^{\mu (j+1)} - 1}$, if $\mu > \ln 2$, and $\rho (\lambda)  = \frac{1}{2^{j+1}-1}$ otherwise.
\end{enumerate}
\end{th4}

\begin{proof}

The first statement is obvious. The spectrum of the binary tree ensemble $\sigma^{\rm ensemble}$ is the union of spectra with every possible $k$. It reads
\be
\sigma^{\rm ensemble} = \bigcup_{k = 1}^{\infty} \sigma(\calT_k) = \bigcup_{k = 1}^{\infty}
\bigcup_{j = 1}^{k}\left\lbrace 2\sqrt{2} \cos \frac{\pi j}{k + 1}\right\rbrace,
\label{eq:33}
\ee
where $\calT_k$ is the full binary tree of level $k$.
Compare to \eq{eq:spec_Tk}, note that $k \rightarrow \infty$ instead of being finite.

Recall that level $k$ is distributed exponentially, or $\PP_{\mu}(k) = (e^{\mu} - 1) e^{-\mu k}$.
The spectral density $\rho (\lambda)$ of the ensemble is the following quotient:
\be
\rho (\lambda) = \frac{\EE_{\mu}\ \tM_k (\lambda)}{\EE_{\mu}\ \text{N}_k} = \frac{\sum\limits_{k =
1}^{\infty} \tM_k(\lambda)\ e^{-\mu k}}{\sum\limits_{k = 1}^{\infty} (2^k - 1)\ e^{-\mu k}}
\ee
Note that both series do not always converge. This fact is due to number of vertices, or tree size, growing exponentially with level. Substituting here the expression for multiplicities,
we get:
\be
\rho \left(2\sqrt2 \cos \frac{\pi i}{j+1}\right) = \dfrac{\dfrac{1}{2^{j+1}-1} \sum\limits_{k =
j+1}^{\infty} (2^k - 2^{k\ \mod (j+1)})\: e^{-\mu k} + \sum\limits_{k = j}^{\infty} e^{-\mu k}\:
\II\{k \equiv j\ \mod (j+1)\}}{\sum\limits_{k=1}^{\infty} (2^k - 1)\: e^{-\mu k}}
\ee
For $\mu>\ln 2 \approx 0.69$ all series converge. Computing the corresponding geometric sums, we obtain
\be
\rho \left(2\sqrt2 \cos \frac{\pi i}{j+1}\right) = \dfrac{(e^{\mu} - 1)^2}{e^{\mu (j+1)} - 1}
\label{eq:38}
\ee
For $\mu<\ln 2$ the sums diverge, however the limit of their partial sums quotient exists, does not
depend on $\mu$ and is equal to
\be
\rho \left(2\sqrt2 \cos \frac{\pi i}{j+1}\right)  = \frac{1}{2^{j+1}-1}
\label{eq:39}
\ee
Also note that this expression is the limit of \eq{eq:38} as $\mu \rightarrow \ln 2$. Another important remark is that these expressions are indeed very similar to the individual case, and this resemblance is especially evident for $\mu < \ln 2$.
\end{proof}

Our analytic results are perfectly consistent with the numeric simulations. Theoretical and practical values are available in tables \ref{tab:3} and \ref{tab:4}.

\begin{table}[ht]
\subfloat[Theoretical peak frequencies for $\mu = 0.3,\ 0.5,\ 0.8$ and $1.0$. The first two columns are calculated from \eq{eq:39}, other two - from \eq{eq:38}.]{ \label{tab:3}
\begin{tabular}{|c||c|c|c|c|}
\hline
Peak No. & $\mu = 0.3$ & $\mu = 0.5$ & $\mu = 0.8$ & $\mu = 1.0$\\
\hline \hline
$0$ & $0.3333$ & $0.3333$ & $0.3799$ & $0.4621$\\
\hline
$1$ & $0.1423$ & $0.1423$ & $0.1498$ &$0.1547$\\
\hline
$2$ & $0.0667$ & $0.0667$ & $0.0638$ & $0.0551$\\
\hline
$3$ & $0.0323$ & $0.0323$ & $0.0280$ & $0.0200$\\
\hline
$4$ & $0.0159$ & $0.0159$ & $0.0125$ & $0.0073$\\
\hline
\end{tabular}
} \hspace{0.5cm} \subfloat[Peak frequencies for the same ensembles according to the simulation.]{ \label{tab:4}
\begin{tabular}{|c||c|c|c|c|}
\hline
Peak No. &$\mu = 0.3$& $\mu=0.5$ & $\mu=0.8$ & $\mu = 1.0$\\
\hline \hline
$0$ & $0.3347$ & $0.3369$ & $0.3882$ & $0.4668$  \\
\hline
$1$ & $0.1433$ & $0.1437$ & $0.1514$ & $0.1548$ \\
\hline
$2$ & $0.0669$ & $0.0667$ & $0.0631$ & $0.0536$\\
\hline
$3$ & $0.0322$ & $0.0320$ & $0.0273$ & $0.0197$\\
\hline
$4$ & $0.0158$ & $0.0156$ & $0.0120$ & $0.0071$\\
\hline
\end{tabular}
}
\caption{Comparison of the theoretical and experimental (computer simulation) values of main peak frequencies in spectral densities of full binary tree ensembles with level distributed exponentially with various $\mu$.}
\end{table}

We are in position to describe the envelope of the spectral density in the ensemble. In this case we again pain attention to ``outer'' and ``inner'' curves. The outer curve $\rho^{out}(\lambda)$ is defined by the series:
\be
\lambda_j^{out} = 2\sqrt2 \cos \frac{\pi}{j+1}, \quad j = 2, 3, 4, \ldots
\ee
while the inner curve $\rho^{in}(\lambda)$ is set by:
\be
\lambda_j^{in} = 2\sqrt2 \cos \frac{\pi}{2j(2j+1)}, \quad j = 2, 3, 4, \ldots .
\ee
Let us formulate the theorem regarding the asymptotics of the inner and outer curves.

\newtheorem{th5}[counter]{Theorem}
\begin{th5} Let $\rho(\lambda)$ be the spectral density of the full binary tree ensemble with level $k$ distributed exponentially with parameter $\mu$, then the outer curve $\rho^{out}(\lambda)$ and the inner curve $\rho^{in}(\lambda)$ have the following asymptotics:
\begin{enumerate}
\item $\rho^{out}(\lambda) \sim A_{\mu} \exp
\left(-\dfrac{B_{\mu}}{\sqrt{\lambda_{max} - \lambda}}\right)$ at $\lambda \rightarrow \lambda_{max}$,
\item $\rho^{in}(\lambda) \sim A_{\mu} \exp \left(-\dfrac{C_{\mu}}{\lambda}\right)$
at $\lambda \rightarrow 0$.
\end{enumerate}
where $A_{\mu}, B_{\mu}, C_{\mu}$ are some positive constants independent of $\lambda$.
\end{th5}

\begin{proof}

First, consider $\mu > \ln 2$. Following the proof of Theorem 3, use the Taylor expansion of $\lambda_j^{out}$ at $j \gg 1$:
\be
\lambda = 2 \sqrt{2}\cos\frac{\pi}{j + 1} = 2\sqrt2\left(1 - \frac{\pi^2}{2 (j+1)^2}\right) +
o\left(\frac{1}{(j+1)^2}\right).
\ee
Inverting this expression, we get
\be
j+1 = \frac{\sqrt[4]{2} \pi}{\sqrt{2\sqrt2 - \lambda}} + o\left(\frac{1}{(j+1)^2}\right)
\ee
and substitute it into the spectral density \eq{eq:38} we obtained in the previous theorem:
\be
\rho(\lambda) \sim (e^{\mu} - 1)^2 \exp \left(-\frac{\sqrt[4]{2} \pi \mu}{\sqrt{\lambda_{max} - \lambda}}\right).
\ee

As for the inner curve, again, use the Taylor expansion of $\lambda_j^{in}$ at $j \gg 1$:
\be
\lambda = 2 \sqrt{2} \sin \frac{\pi}{2(2j + 1)} = \frac{\pi}{\sqrt2 (2j+1)} +
o\left(\frac{1}{2j+1}\right),
\ee
then invert it, getting
\be
2j + 1 = \frac{\pi}{\sqrt{2} \lambda} + o\left(\frac{1}{2j+1}\right).
\ee
This expression then is substituted into \eq{eq:38} and the final result is
\be
\rho(\lambda) \sim (e^{\mu} - 1)^2 \exp \left(\-\frac{\sqrt2 \pi \mu}{\lambda}\right).
\ee

Collecting these two formulas together, we get the shape of $\rho(\lambda)$:
\be
\rho (\lambda) \sim \left\{\begin{array}{ll}A_{\mu}\ \exp
\left(-\dfrac{B_{\mu}}{\sqrt{\lambda_{max} - \lambda}}\right), & \mbox{for $\lambda\to\lambda_{max}$}
\medskip \\ A_{\mu}\ \exp \left(-\dfrac{C_{\mu}}{\lambda}\right), & \mbox{for $\lambda\to 0$}
\end{array} \right.
\label{eq:40}
\ee
where $A_{\mu} = (e^{\mu} - 1)^2$, $B_{\mu} = \sqrt[4]{2} \pi \mu$ and $C_{\mu} = \sqrt2 \pi \mu$. Recall that the density in the case of $\mu < \ln 2$ is simply the same as if we just substituted $\mu = \ln 2$ in the case of $\mu > \ln 2$.
It means that to obtain the expressions of the inner and outer curves we also have to assume $\mu = \ln 2$ in \eq{eq:40}. By doing that we get exactly the same expressions as \eq{eq:30} and \eq{eq:32}, corresponding to $A_{\mu} = 1,\ B_{\mu} = \sqrt[4]{2} \pi \ln 2$ and $C_{\mu} = \sqrt2 \pi \ln 2$.
\end{proof}

To sum it up, we have evaluated the spectra of a single full binary tree and of the exponentially weighted ensemble of full binary trees by computing explicitly all multiplicities (frequencies). We found $\rho(\lambda)$ at the edges of the spectrum, known as the ``Lifshitz tail'' \cite{lifshitz,pastur}, typical for the Anderson localization in disordered one-dimensional systems such as the one-dimensional Schr\"odinger-like discrete operators with a diagonal disorder. Two interesting peculiarities of considered systems are worth to mention: i) even a single large tree has the Lifshitz tail; ii) the spectral density of the exponentially weighted ensemble of full binary trees has the same asymptotic behavior at the edge of the spectrum as the exponential ensemble of linear chains.

\subsection{Star graphs: numerics}

\begin{figure}[ht]
\subfloat[Spectral density of
$\calS_{6,3}$]{\includegraphics[width=0.45\textwidth]{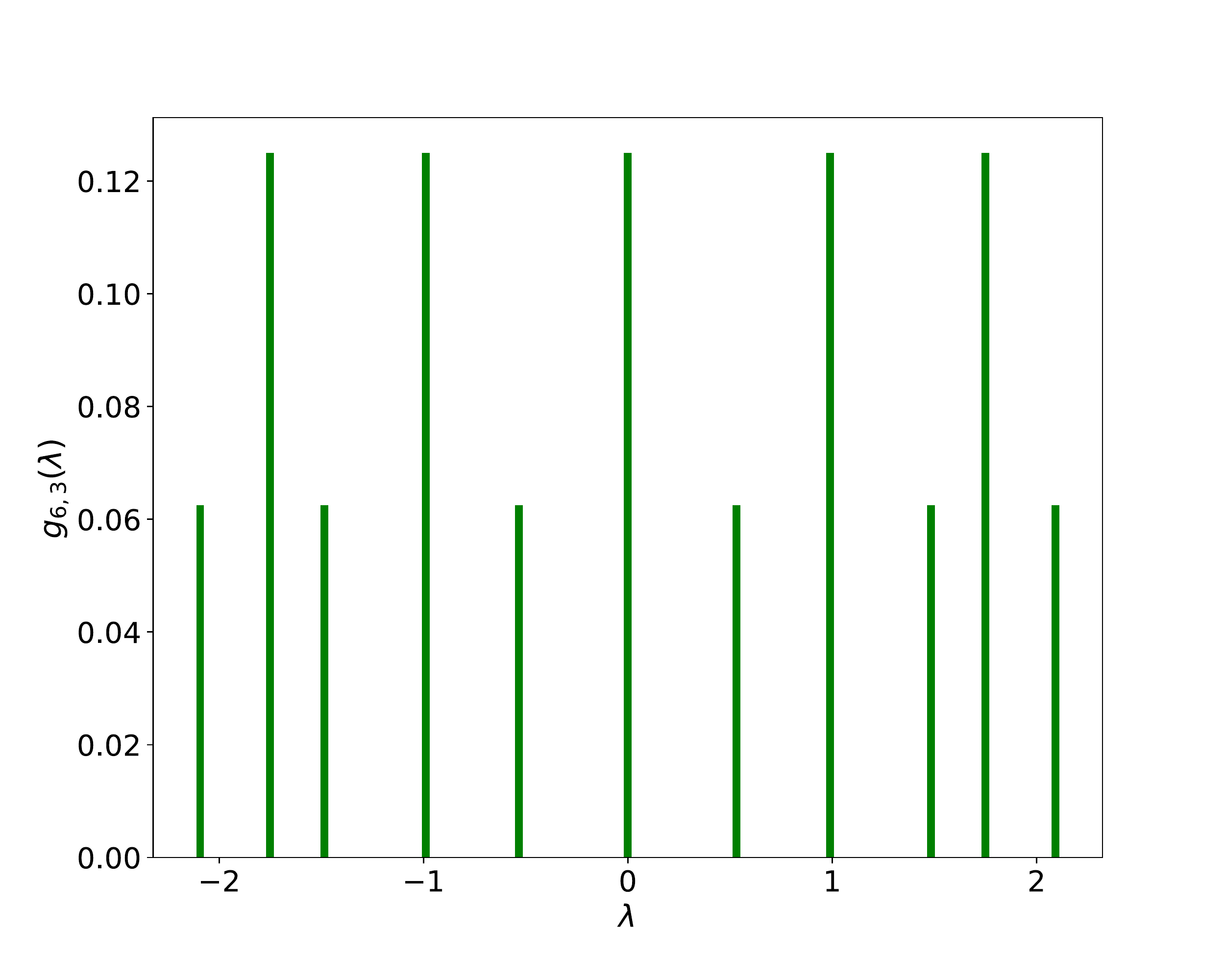}} \subfloat[Spectral density
of $\calS_{11,4}$]{\includegraphics[width=0.45\textwidth]{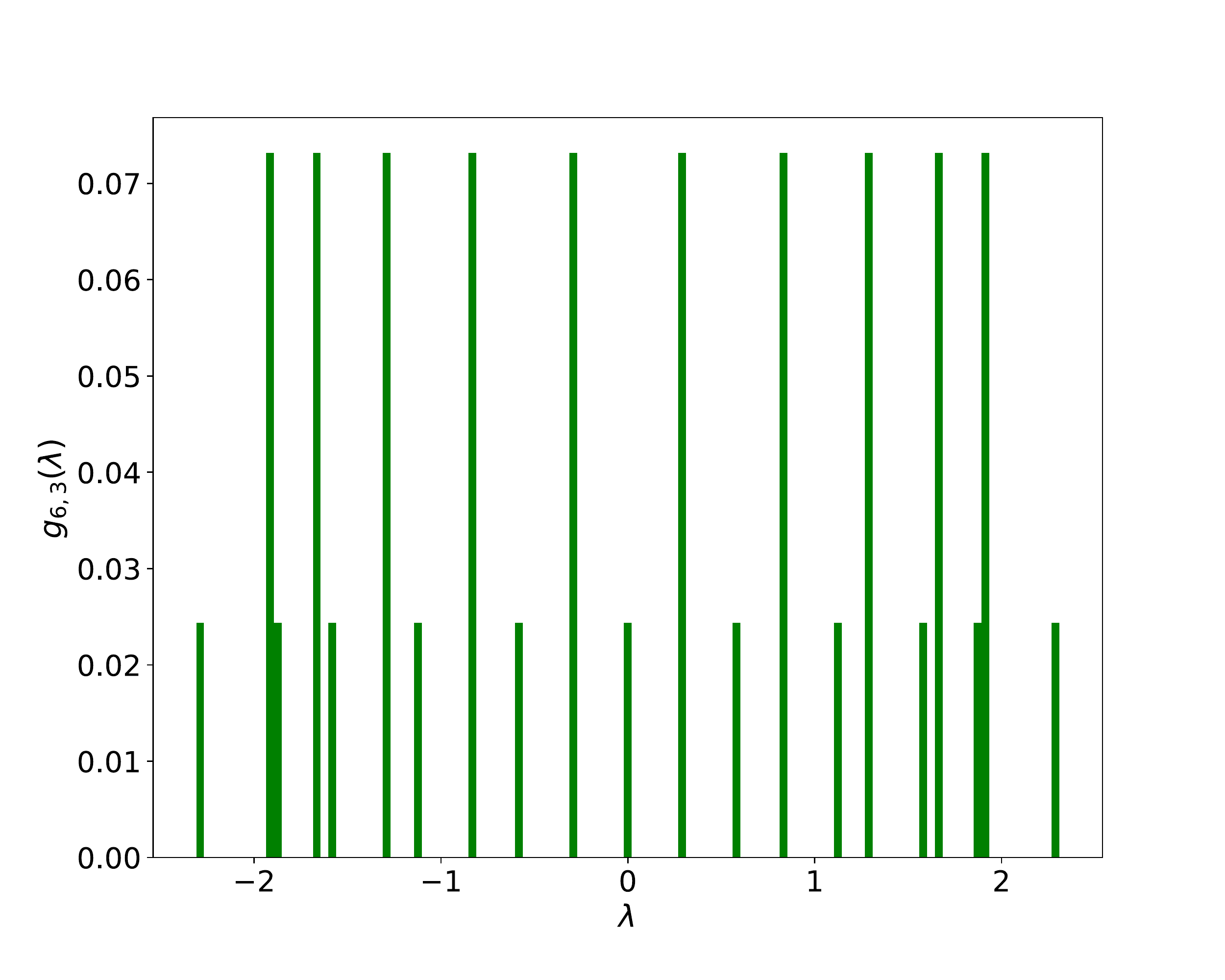}}
\caption{Eigenvalue distribution for stars. Axis X is eigenvalue, axis Y is its frequency.}
\label{fig:st_single}
\end{figure}

We begin the analysis of the star-like graphs by calculating their spectra with different parameters $k$ and $p$ (see Fig. \ref{fig:trees1}b). In particular, we consider $k = 1, \ldots,
500$ and $p = 3, \ldots, 10$, which is possible as the size of a star is linear in terms of both level $k$ and root degree $p$. We emphasize that the spectrum and spectral density of an individual star do depend on $k$ and $p$, but their structure is fixed. The spectrum always consists of two separate series: $k-1$ eigenvalues of multiplicity $p-1$ and $k$ simple eigenvalues as one can observe in \fig{fig:st_single}. Note that almost all eigenvalues belong to the segment $[-2; 2]$, and there are also two symmetric with respect to zero outstanding eigenvalues.

Unlike the case of full binary trees, the spectral density of the exponential star ensemble is essentially different from the individual star spectral density. What is more, the structure of this spectral density depends on $\mu$ notedly as shown in \fig{fig:st_ens}. This behavior should be compared to the one considered in \cite{root} for tree-like graphs with a ``heavy'' root, where the localization-delocalization of the spectrum has been found.

\subsection{Star graphs: theory}

\subsubsection{Spectrum of a single star graph}

Here we derive the spectrum of an individual star graph. We pay special attention of computing multiplicities (frequencies) of eigenvalues.

\newtheorem{th6}[counter]{Theorem}
\begin{th6} Let $\calS_{k,p}$ be the star of $k$ levels and root degree $p$, $\sigma(\calS_{k,p})$ - its spectrum and $g_{k,p} (\lambda)$ - its spectral density, then:
\begin{enumerate}
\item $\sigma(\calS_{k,p}) =  \sigma_k^{lin} \cup \sigma_{k,p}^c$, where $\sigma_k^{lin} = \bigcup\limits_{j = 1}^{k-1}\left\lbrace 2
\cos \dfrac{\pi j}{k}\right\rbrace$ and $\sigma_{k,p}^c$ is the set of root of the complex equation $ \tan k\varphi~=~\frac{p}{p-2} \tan \varphi,\ \lambda = 2 \cos \varphi$.
\item $g_{k,p}(\lambda) =
\begin{cases}
\dfrac{p-1}{p\ (k-1) + 1}, & \mbox{if $\lambda \in \sigma_k^{lin}$}, \medskip \\
\dfrac{1}{p\ (k-1) + 1}, & \mbox{if $\lambda \in \sigma^c_{k,p}$}
\end{cases}$
\end{enumerate}
\end{th6}

\begin{proof}

For the star $\calS_{k,p}$ of $p$ branches and $k$ levels, shown in \fig{fig:trees1}b,
according to Theorem 1 \cite{rojo2007explicit}, the spectrum is the set of eigenvalues of $j \times
j$ principal submatrices $(j = 1 \ldots k)$.
\be
\bA(\calS_{k,p})=
\begin{bmatrix}
0 & 1 && \\
1 & 0 & 1 && \\
&\hdotsfor{4} & \\
&&& 1 & 0 & \sqrt{p} \\
&&&& \sqrt{p} & 0
\end{bmatrix}
\ee
First, Theorem 1 suggest constructing the set $\Omega$ of indices such that $n_j > n_{j+1}$ as the proposed multiplicities are
\be
m_j = n_j - n_{j+1},\; j = 1 \ldots k - 1;\ m_k = 1.
\label{eq:41}
\ee
We notice that this set consists of one index only as for star graphs $n_j = p$ for $j = 1, \ldots, k - 1$, and $n_k = 1$. Hence, $m_j = 0$ for $j = 1, \ldots, k-2$, and we only need to compute the spectra of $\bA_{k-1}$ (eigenvalues of which have multiplicity $p-1$) and $\bA_k$ (eigenvalues of which are simple). These are the two separate series we pointed out in the numerical section.

This leads us to the following conjecture: we split the spectrum of a star into two parts. The first one is the linear chain contribution $\sigma_k^{lin}$ as the spectrum of $\bA_{k-1}$, and the second one in the center contribution $\sigma_{k,p}^c$ as the spectrum of $\bA_{k}$. As the total number of eigenvalues equals $p(k-1) + 1$, the final expression for the spectral density as in \eq{eq:04} is
\be
g_{k,p} (\lambda) = \frac{\tM_{k,p} (\lambda)}{p(k-1) + 1} =
\begin{cases}
\dfrac{p-1}{p\ (k-1) + 1}, & \mbox{if $\lambda \in \sigma_k^{lin}$}, \medskip \\
\dfrac{1}{p\ (k-1) + 1}, & \mbox{if $\lambda \in \sigma^c_{k,p}$}
\end{cases}
\ee

The roots of $F_{k-1}$, derived in \eq{eq:13}, define the spectrum $\sigma_k^{lin}$:
\be
\sigma_k^{lin} = \bigcup_{j = 1}^{k-1}\left\lbrace 2 \cos \frac{\pi j}{k}\right\rbrace
\label{eq:42}
\ee

Computing $\sigma_{k,p}^c$ is a much more difficult task. Applying the Laplace formula to the last row of matrix $\bA_k$ two times, we get:
\be
G_k = \det \bA_k = \lambda F_{k-1} - p F_{k-2} = \frac{p\mu_+ - \lambda}{ \sqrt{\lambda^2 - 4}}
\mu_-^k - \frac{p\mu_- - \lambda}{ \sqrt{\lambda^2 - 4}} \mu_+^k
\label{eq:43}
\ee
where
\be
\mu_{\pm}= \frac{\lambda \pm \sqrt{\lambda^2 - 4}}{2}
\label{eq:44}
\ee
The equation $G_k = 0$ is equivalent to the following complex-valued transcendental equation:
\be
\tan k\varphi = \frac{p}{p-2} \tan \varphi,
\label{eq:45}
\ee
where we have introduced $\varphi$ as follows:
\be
\tan \varphi = \frac{\sqrt{4 - \lambda^2}}{\lambda}, \quad (\lambda = 2 \cos \varphi).
\label{eq:46}
\ee
The roots of this equation form $\sigma_{n,p}^c$. This set is always unique, except for
the roots of type $\lambda = 2 \cos \frac{\pi i}{j}$, where $i$ and $j$ are coprime, or in other
words, for any $n \neq l$:
\be
(\sigma_{n,p}^c \cap \sigma_{l,p}^c) \setminus \sigma^{lin} = \varnothing
\label{eq:47}
\ee
And also for any $n$
\be
\sigma_{n,p}^c \cap \sigma_n^{lin} = \varnothing
\label{eq:48}
\ee
\end{proof}

\newtheorem{th7}[counter]{Theorem}

\begin{th7} Let $\calS_{k,p}$ be the star with $p$ branches and $k$ levels, $\sigma(\calS_{k,p})$ be its spectrum and $\sigma_{k,p}^c$ be the central contribution in $\sigma(\calS_{k,p})$, then
\begin{enumerate}
\item the maximal eigenvalue in the whole spectrum belongs to $\sigma_{k,p}^c$ and lies in the interval $(\sqrt{p}; \frac{p}{\sqrt{p-1}})$.
\item the rest of the values from $\sigma_{k,p}^c$ belong to the interval $(-2;2)$ and each positive value localizes in the interval $\lambda_t \in \left(2 \cos \left(\frac{\pi t}{k} + \frac{1}{k} \arctan \left(\frac{p}{p-2} \tan \frac{\pi t}{k}\right)\right); 2\cos \left(\frac{\pi t}{k} + \frac{\pi}{2k}\right)\right)$, $t = 1, 2, \ldots, \left[\frac{k-2}{2}\right]$, with a corresponding negative value of the same absolute value. There is also $\lambda = 0$ for any odd $k$.
\end{enumerate}
\end{th7}
\begin{proof}
Recall that $\sigma_{k,p}^c$ is the set of roots of the complex-valued equation
\be
\tan k\varphi = \frac{p}{p-2} \tan \varphi \quad (\lambda = 2 \cos \varphi).
\ee
The maximal eigenvalue $\lambda_k^{max}$ is always the eigenvalue of $\bA_k$, so it always belongs to $\sigma_{k,p}^c$.

Despite this equation cannot be solved exactly, a transparent analysis of its solution is available. Here is a brief explanation of how we localize the maximal root, $\lambda^{max}_k$ for various $p$ and $k$. Computing $G_k$ as defined in \eq{eq:43} for small $k$, we get ($p\ge 3$)
\be
\begin{array}{lll}
G_2 = \lambda^2 - p & \Rightarrow & \lambda_2^{max} = \sqrt{p}, \medskip \\
G_3 = \lambda (\lambda^2 - (p+1)) & \Rightarrow & \lambda_3^{max} = \sqrt{p+1} \medskip \\ & ...
\end{array}
\ee
For $\lambda \ge 2$ the equation \eq{eq:46} transforms into the following real-valued equation:
\be
\tanh \varphi = \frac{\sqrt{\lambda^2 - 4}}{\lambda}\quad (\lambda = 2 \cosh \varphi).
\ee
For $\varphi > 0$ the equation \eq{eq:45} has either no roots, or the only root since the functions $y_k
(\varphi) = \tanh k\varphi$ and $z (\varphi)~=~\frac{p}{p-2} \tanh \varphi$ are both monotone and
concave. The illustration of this can be seen in \fig{fig:tanh}. If $y_k^{\prime}(0)  \le z^{\prime}(0)$, there are no roots, and there is one otherwise, which happens when $k>\frac{p}{p-2}$.

Note that $p\ge 3$, so $\frac{p}{p-2} \le 3$. Also note that if $k > l$, $y_k > y_l$ for any
$\varphi > 0$. As a result, $\lambda_k^{max}$ is strictly monotonic and $\lambda_2^{max} =
\sqrt{p}$ is the minimal value in this series. For any $k$ the function $y_k(\varphi) < 1$, thus
the upper bound for the solution is:
\be
\frac{p}{p-2} \tanh \varphi = 1 \quad \Rightarrow \quad \lambda = \frac{p}{\sqrt{p - 1}}
\ee

\begin{figure}[ht]
\centering \subfloat[Plots for $p=3$] {\includegraphics[width=0.45\textwidth]{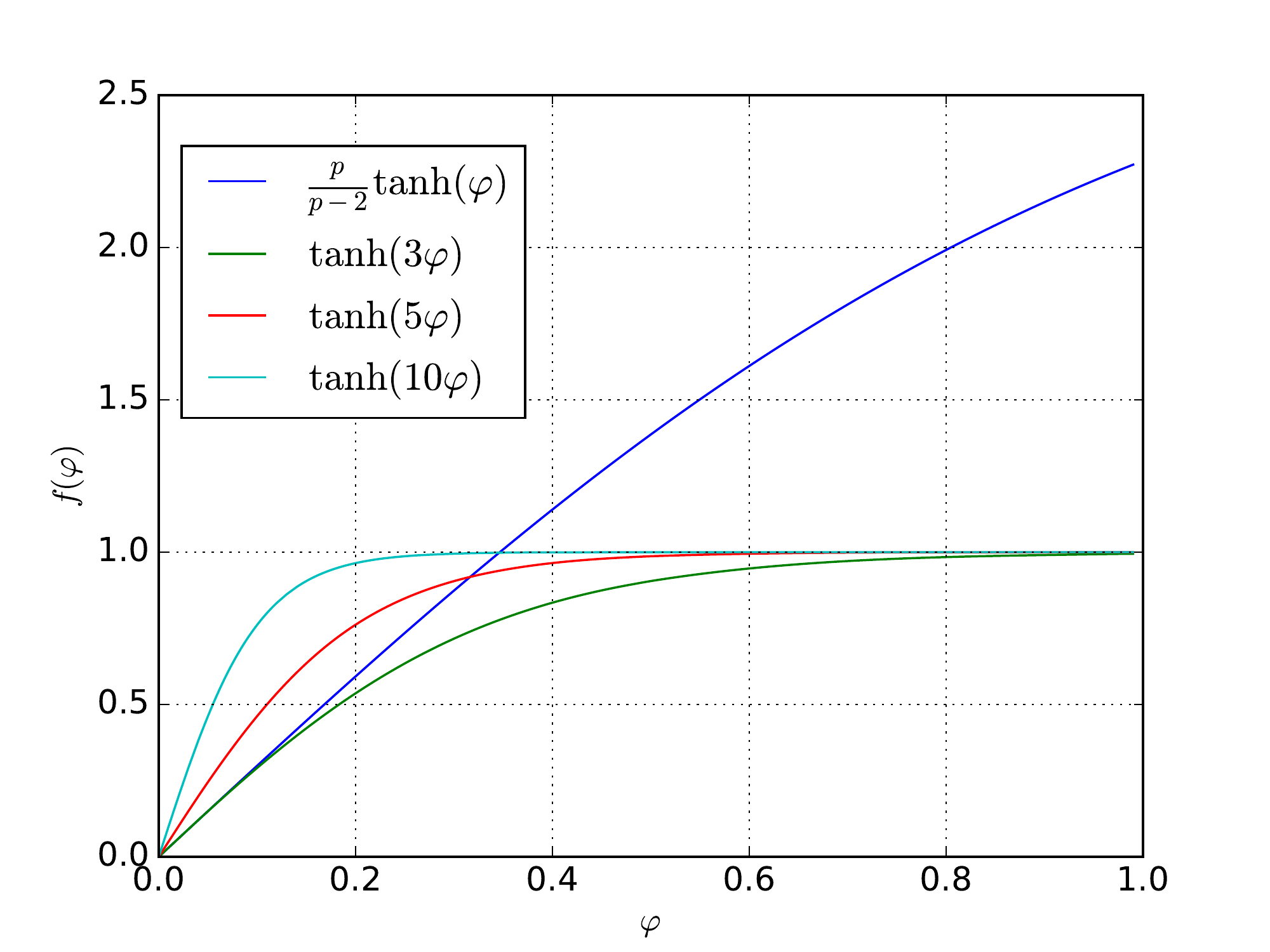}}
\subfloat[Plots for $p=4$]{\includegraphics[width=0.45\textwidth]{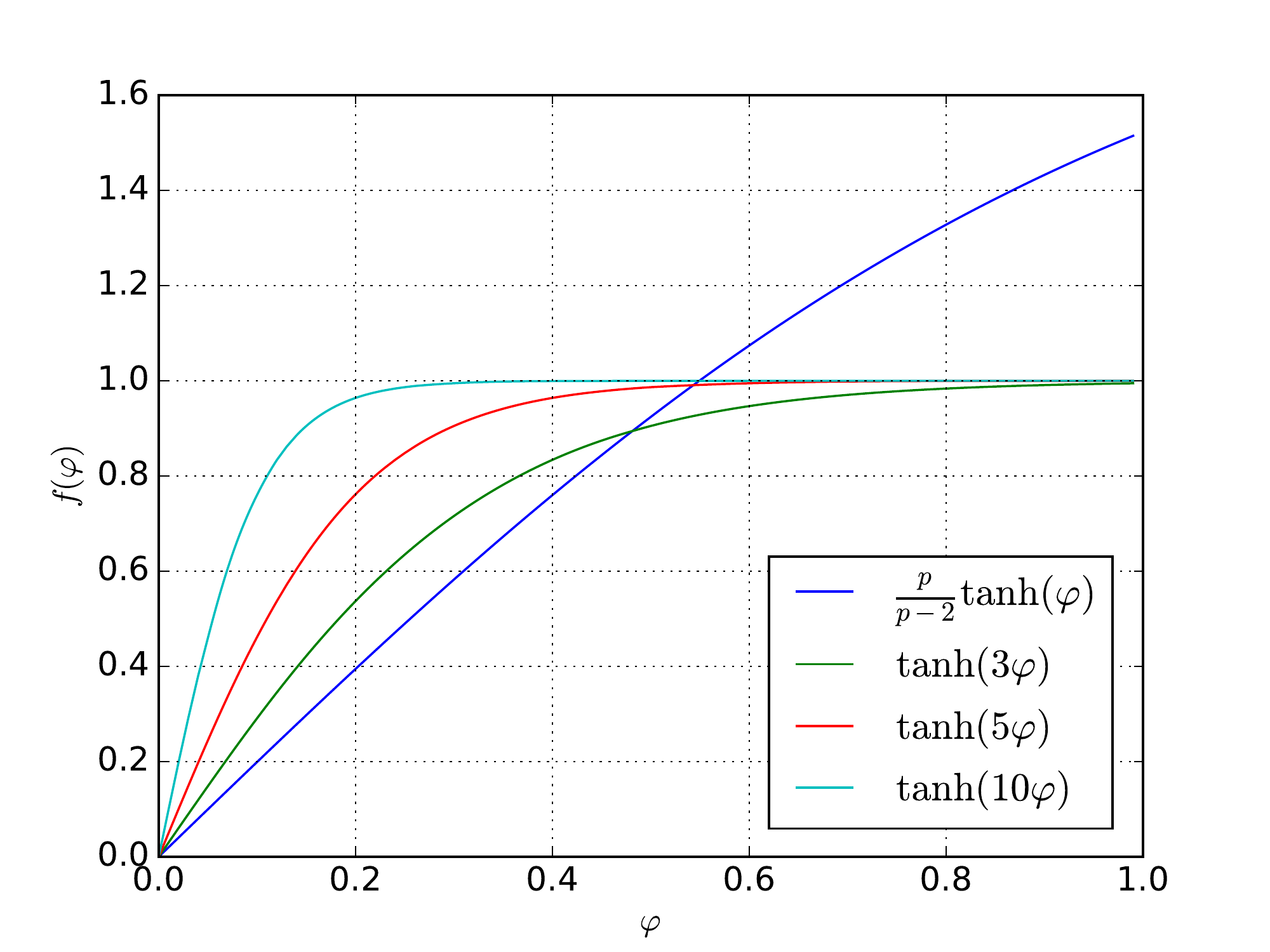}}
\caption{Illustration of graphic solution of \eq{eq:45}.}
\label{fig:tanh}
\end{figure}

So, we proved that $\lambda_k^{\max} \in (\sqrt{p}, \frac{p}{\sqrt{p-1}})$. Now we are moving on to the second statement of the theorem.

For $\lambda \in (-2;2)$ the equation \eq{eq:46} transforms into the real-valued equation in the same form, but with real $\varphi$
\be
\tan k\varphi = \frac{p}{p-2} \tan \varphi .
\ee
First, let us count its roots. It is clear that we should only count the roots in the segment $[0; \frac{\pi}{2}]$ as the number of negative roots is the same as of positive ones and due to both functions being periodic, all other values would give the same $\lambda = 2\cos \varphi$.

The period of $f_k(\varphi) = \tan k\varphi$ is $\frac{\pi}{k}$, the period of $f_0(\varphi) = \frac{p}{p-2} \tan \varphi$ is $\pi$. There are no non-zero roots in the segment $[0; \frac{\pi}{2k}]$ as the derivative of $f_k$ is always larger than that of $f_0$. Zero is excluded from the possible roots by the denominator in $G_k$. In every other period of $f_k$, the segment $[\frac{\pi}{2k} + \frac{\pi t}{k}; \frac{\pi}{2k} + \frac{\pi (t+1)}{k}]$ with $t = 1,2, \ldots, \left[\frac{k}{2}\right]$, there is strictly one root, as $f_k$ is unbounded and $f_0$ is bounded and both are convex and monotone.

If $k$ is even, the last period of $f_k$ is actually a semi-period, so there are no roots there belonging to $[0; \frac{\pi}{2}]$. Thus there are $\frac{k}{2} - 1$ roots overall, which makes it $k-2$ in the segment $[-\frac{\pi}{2}; \frac{\pi}{2}]$  and gives us $k-2$ different $\lambda$. If $k$ is odd, we consider $\varphi = \frac{\pi}{2}$ a solution as well, which corresponds to $\lambda = 0$. So, this gives us $\frac{k-1}{2} \cdot 2 - 1$ different $\lambda$, which is again $k-2$. This idea is illustrated by \fig{fig:tan}. Recall that there were two more values. So, that makes $k$ roots altogether, which is exactly the number of roots for a polynomial of degree $k$. This means that we have found all the roots: $k-2$ in the interval $(-2;2)$ and two more in $(-\frac{p}{\sqrt{p-1}}; -\sqrt{p}) \cup (\sqrt{p}, \frac{p}{\sqrt{p-1}})$.

Now consider $k \rightarrow + \infty$. Recall that we indexed the periods of $f_k$ with $t$. For small $t$ every period $f_0$ is almost constant. This means that the solution of $f_0(\varphi) = f_k(\varphi)$ approximately suffices
\be
f_k(\varphi) = \tan k \varphi = \frac{p}{p-2} \tan \frac{\pi t}{k} \approx f_0\left(\frac{\pi t }{k}\right).
\ee
This would actually give the lower bound on the $t$-th root $\varphi_t$. So, the approximate solution lies in the interval
\be
\varphi_t \in \left(\frac{\pi t}{k} + \frac{1}{k} \arctan \left(\frac{p}{p-2} \tan \frac{\pi t}{k}\right); \frac{\pi t}{k} + \frac{\pi}{2k}\right).
\ee
For bigger $t$ the root tends to the right end of the segment, which is $\frac{\pi}{2k} + \frac{\pi t}{k}$. The corresponding $\lambda$ then lies in the following interval:
\be
\lambda_t \in \left(2 \cos \left(\frac{\pi t}{k} + \frac{1}{k} \arctan \left(\frac{p}{p-2} \tan \frac{\pi t}{k}\right)\right); 2\cos \left(\frac{\pi t}{k} + \frac{\pi}{2k}\right)\right),
\ee
where $t$ is the index of the positive root counting from the right edge of the spectrum.
\end{proof}

\begin{figure}[ht]
\centering \subfloat[Plots for $p=3,4,5$ and $k = 10$.] {\includegraphics[width=0.45\textwidth]{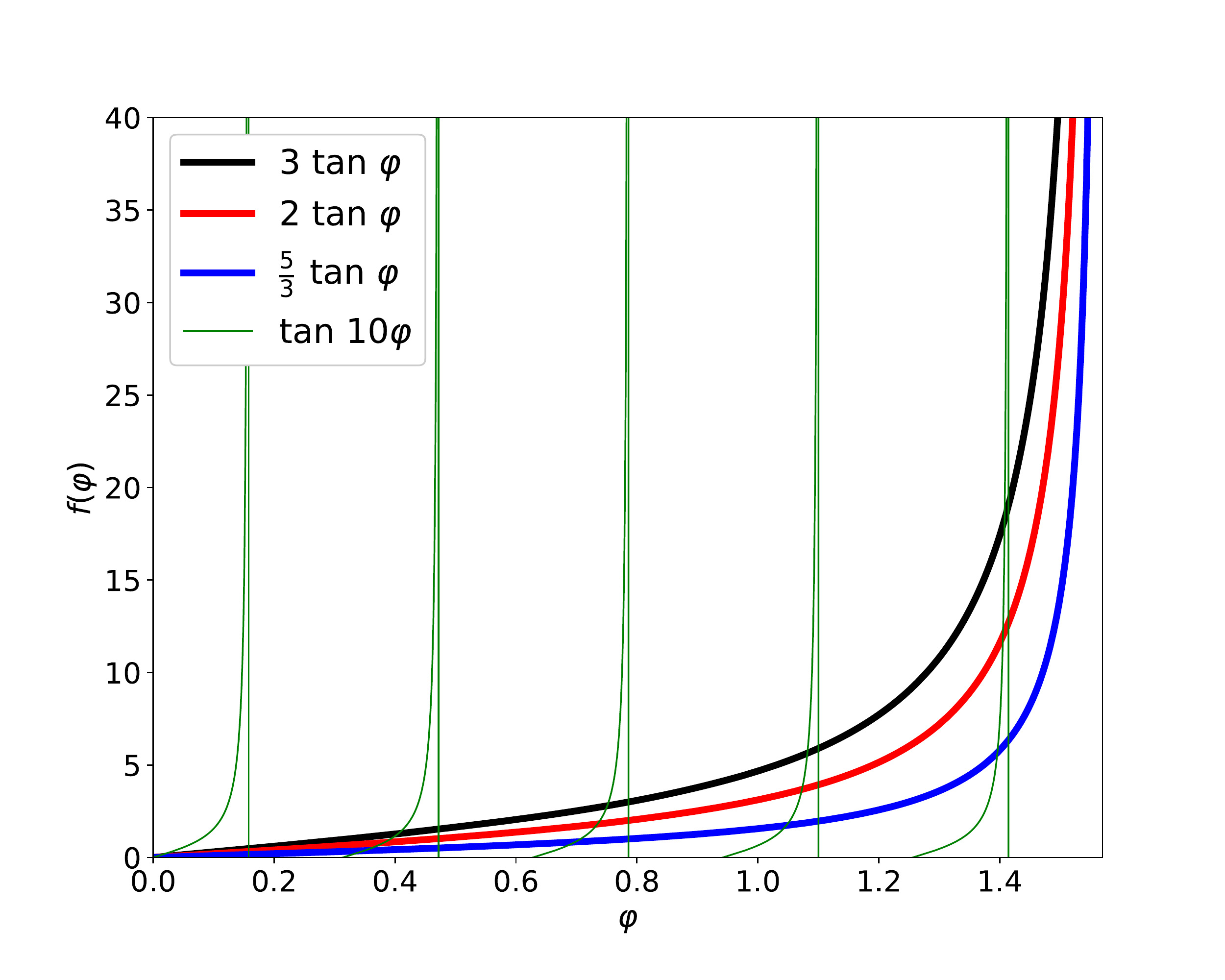}}
\subfloat[Plots for $p=3,4,5$ and $k = 15$.]{\includegraphics[width=0.45\textwidth]{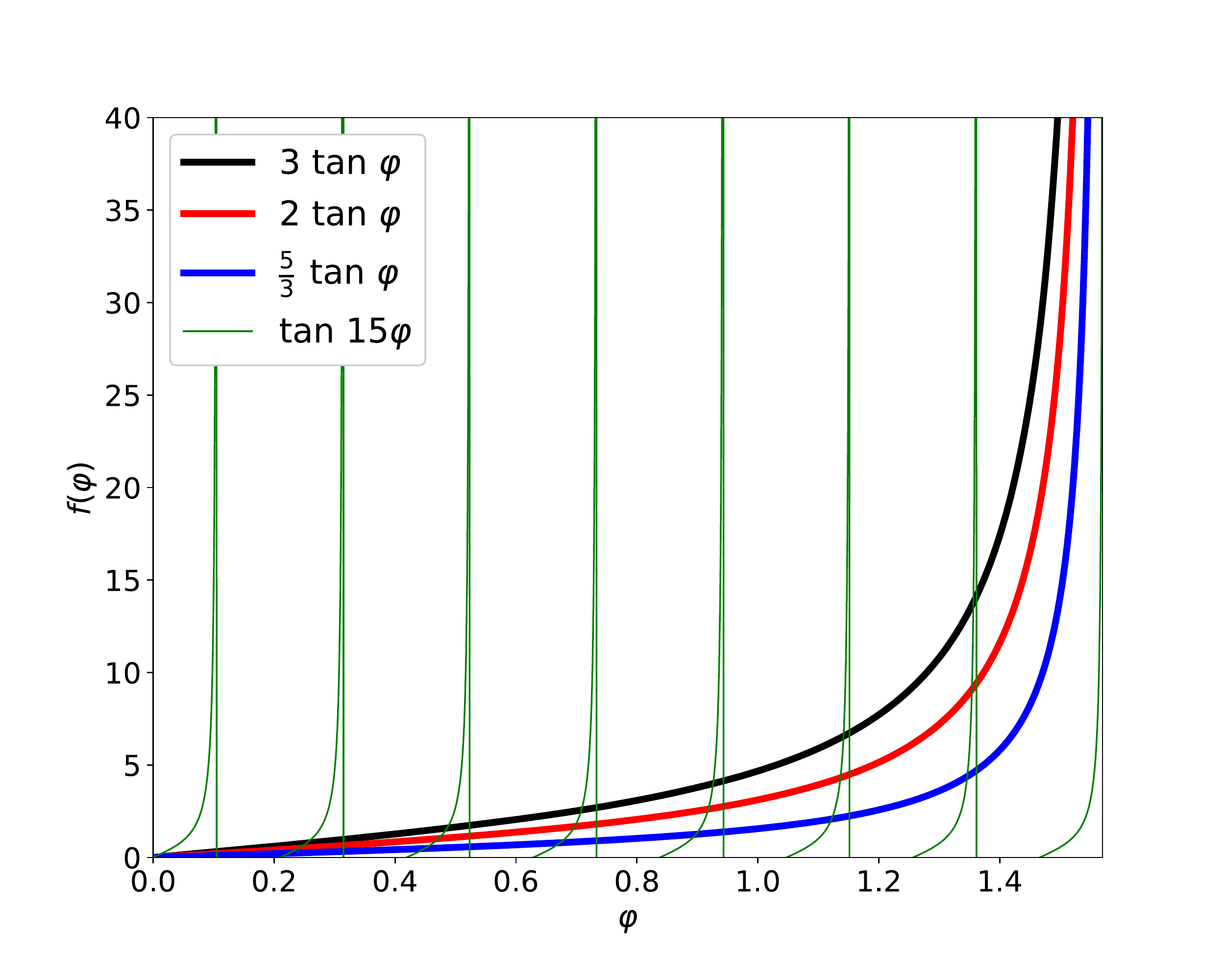}}
\caption{Illustration of the root localization for the equation $\tan k\varphi = \frac{p}{p-2} \tan \varphi $}
\label{fig:tan}
\end{figure}

Recall \fig{fig:st_single} and notice that, indeed the values from $\sigma_{k,p}^c$ tend to the values from $\sigma_p^{lin}$ near the edge.

\subsubsection{Spectral density of ensemble of star graphs}

In this section we derive the spectral density of the star graph ensemble. We also discuss its connection to the linear chain ensemble. Here we again discriminate the two parts of the spectrum:
\be
\sigma_p^{ensemble} =  \sigma^{lin} \cup \sigma_p^{c},
\ee
where $p$ is the degree of the root, and
\be
\sigma^{lin} = \bigcup_{k=2}^{+\infty} \sigma_k^{lin},\quad
\sigma_p^{c} = \bigcup_{k=1}^{+\infty} \sigma_{k,p}^{c}.
\ee
Note that, formally, $\sigma(\calS_{1,p}) = \sigma_{1,p}^{c}$ and $\sigma_1^{lin} = \varnothing$. For this purpose we also divide the spectral density into the sum of contributions from the ``arms'' (linear chain contribution) and the ``root'' (center contribution),
\be
g_p (\lambda) = g_p^{lin} (\lambda) + g_p^{c} (\lambda)
\ee
and formulation the following theorem using this notation.

\newtheorem{th8}[counter]{Theorem}
\begin{th8} Let $\sigma_p^{ensemble}$ be the spectrum of the star ensemble with level distributed exponentially with parameter $\mu$ and $g_p (\lambda) = g_p^{lin} (\lambda) + g_p^{c} (\lambda)$ be the spectral density of such ensemble, then
\begin{enumerate}
\item $\sigma_p^{ensemble} = \bigcup_{k = 2}^{\infty}
\bigcup_{j = 1}^{k-1}\left\lbrace 2 \cos \frac{\pi j}{k}\right\rbrace \cup \bigcup_{k = 1}^{\infty} \sigma^c_{k,p}$,
\item $g_p^{lin}(\lambda) = \frac{(p-1)(e^{\mu} - 1)^2}{(e^{\mu j} - 1) (e^{\mu} + p - 1)}$,
\item $g_p^c (\lambda) = \frac{e^{-\mu k}\ (e^{\mu}-1)^2}{e^{\mu} + p - 1}$.
\end{enumerate}
\end{th8}

\begin{figure}[ht]
\centering \subfloat[$\mu=0.1$] {\includegraphics[width=0.3\textwidth]{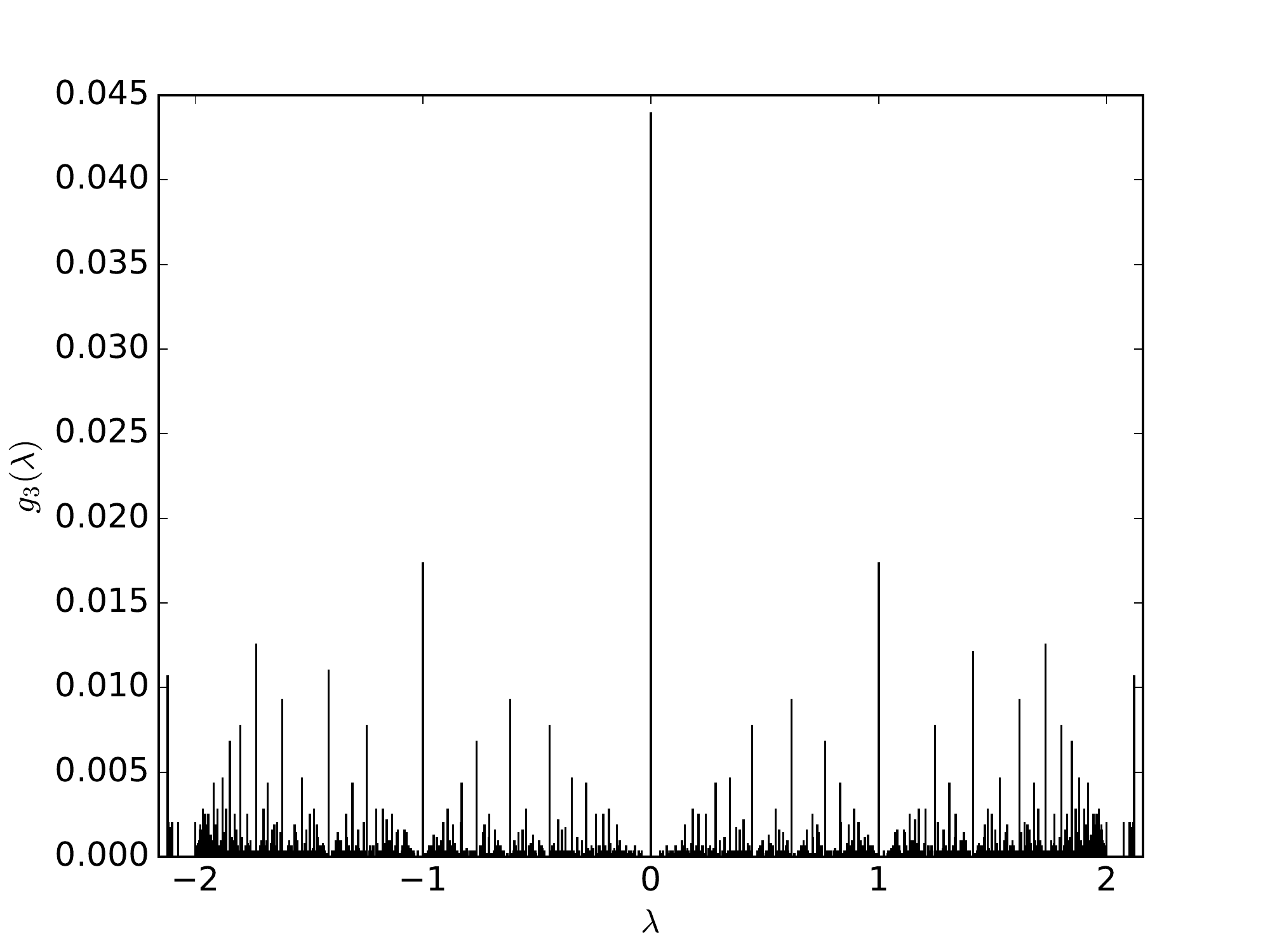}}
\subfloat[$\mu=0.5$]{\includegraphics[width=0.3\textwidth]{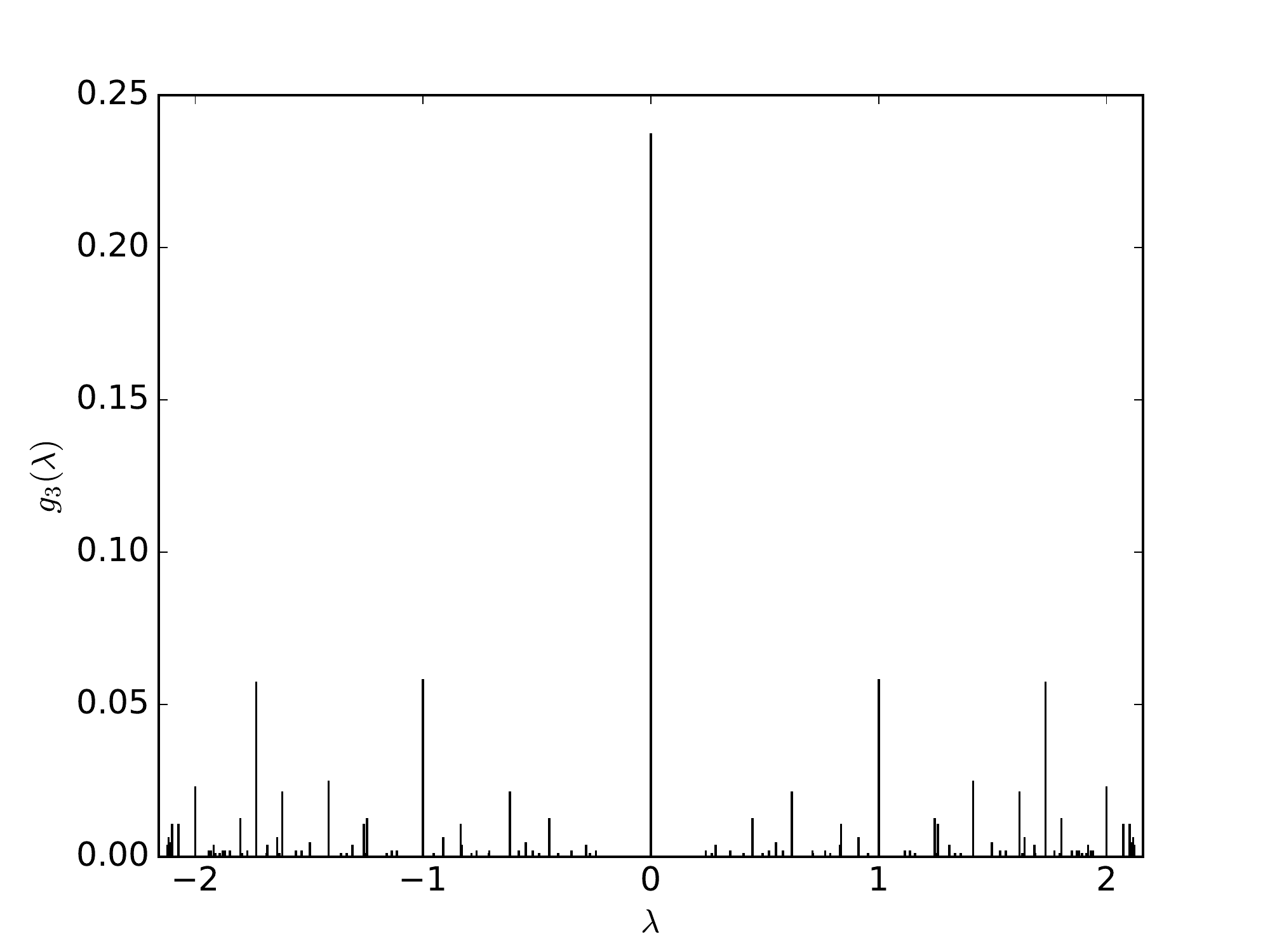}}
\subfloat[$\mu=1.0$]{\includegraphics[width=0.3\textwidth]{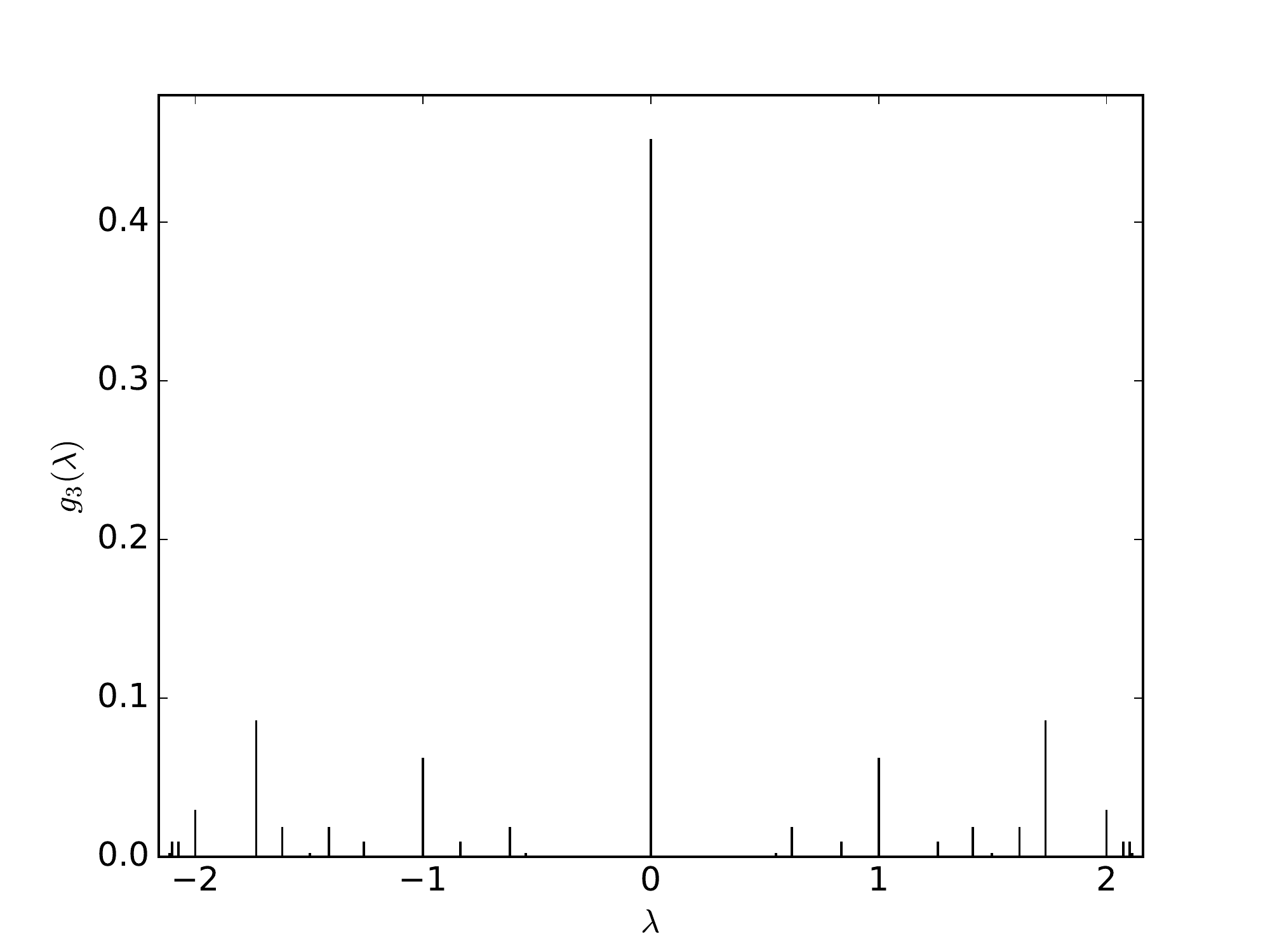}}
\caption{Eigenvalue distribution of star-like graphs with root branching $p=3$ in exponential
ensembles with various $\mu$. Axis X is eigenvalue, axis Y is its frequency. }
\label{fig:st_ens}
\end{figure}

\begin{proof}

The first statement is trivial:
\be
\sigma^{lin} = \bigcup_{k = 2}^{\infty} \sigma_k ^{lin} = \bigcup_{k = 2}^{\infty}
\bigcup_{j = 1}^{k-1}\left\lbrace 2 \cos \frac{\pi j}{k}\right\rbrace, \quad \sigma_p^c =
\bigcup_{k = 1}^{\infty} \sigma^c_{k,p}.
\ee

Any $\lambda \in \sigma^{lin}$ can be written in the form $\lambda = 2 \cos \frac{\pi i}{j}$, where
$i$ and $j \in \NN$ and are coprime. Such $\lambda$ belongs to $\sigma_{tj}^{lin}$ for any $t \in
\NN$ with the multiplicity $p-1$. Then the spectral density $g_p^{lin}(\lambda)$ reads:
\be
g_p^{lin} (\lambda) = \frac{\sum\limits_{k = 1}^{\infty} e^{-\mu k} (p - 1)\ \II\{\lambda \in
\sigma_k^{lin}\}}{\sum\limits_{k=1}^{\infty} (p(k-1) + 1)\ e^{-\mu k}} = \frac{(p-1)\
\sum\limits_{k = 1}^{\infty} e^{-\mu jk}}{\sum\limits_{k=1}^{\infty} (p(k-1) + 1)\ e^{-\mu k}}
\label{eq:53}
\ee
Both series, in the nominator and in the denominator of \eq{eq:53}, converge for any $\mu>0$.
Evaluating these series, we get:
\be
g_p^{lin}(\lambda) = \frac{(p-1)(e^{\mu} - 1)^2}{(e^{\mu j} - 1) (e^{\mu} + p - 1)}
\label{eq:54}
\ee

The function $g_p^{lin}(\lambda)$ is the rescaled spectral density of the linear chain ensemble,
derived in \cite{avetisov2015native}:
\be
g_p^{lin}(\lambda) = \frac{(p - 1) e^{-\mu}}{1 + (p - 1) e^{-\mu}} \rho_{lin}(\lambda) =
\frac{(p - 1) e^{-\mu}}{1 + (p - 1) e^{-\mu}}\  \lim_{\substack{N \rightarrow \infty \\
\varepsilon \rightarrow 0}} \frac{\varepsilon}{\pi N} \sum_{n = 1}^{N} \mu^n \sum_{k=1}^{n}
\frac{1}{(\lambda - 2 \cos \frac{\pi k}{n+1})^2 + \varepsilon^2}
\label{eq:gplin}
\ee
Hence, $g_p^{lin}(\lambda)$ shares all the properties of $\rho_{lin}(\lambda)$.

For a unique eigenvalue $\lambda$ from $\sigma_p^c$, such that $\lambda \in \sigma_{k,p}^c$ one has:
\be
g_p^c (\lambda) = \frac{e^{-\mu k}}{\sum\limits_{k=1}^{\infty} (p(k-1) + 1)\ e^{-\mu k}}=
\frac{e^{-\mu k}\ (e^{\mu}-1)^2}{e^{\mu} + p - 1}
\label{eq:55}
\ee
Most of the values from $\sigma_p^{c}$ are unique as they generally do not satisfy $\lambda = 2
\cos \frac{\pi i}{j}$. We should also point out that $\lambda = 0$ belongs to $\sigma_{k,p}^c$ for
any odd $k$ and to $\sigma_k^{lin}$ for any even $k$:
\be
g_p^c(0) = \frac{\sum\limits_{k = 1}^{\infty} e^{-\mu (2k - 1)}}{\sum\limits_{k=1}^{\infty} (p(k-1)
+ 1) e^{-\mu k}} = \frac{(e^{\mu} - 1)^2}{(e^{2\mu} - 1)(1 + (p-1)e^{-\mu})}
\ee
and
\be
g_p^{lin} (0) = \frac{(p-1)(e^{\mu} - 1)^2}{(e^{2\mu} - 1) (e^{\mu} + p - 1)}
\ee
Thus, we get that frequency (degeneracy) of the central peak does not depend on $p$:
\be
g_p (0) = g_p^{lin} (0) + g_p^c(0) = \frac{e^{\mu} - 1}{e^{\mu} + 1}
\ee
\end{proof}
This result is perfectly consistent with numeric simulations. Another notable result is that \eq{eq:gplin} shows that the connection we assumed does exist. This effect is illustrated by \fig{fig:st_ens}.

We should point out that there is no Lifshitz tail for the enveloping curve like the one found in \eq{eq:40}. However, still for ensemble of stars we have a kind of a Lifshitz tail in the main part of the spectrum near $\lambda$ = 2, just the same as for linear chains \cite{avetisov2015native}. The ``hand-waving'' explanation is as follows: the height of the linear part of the spectrum is $p-1$ times bigger and each value $2\cos \frac{\pi}{k}$ occurs in every $k$-th level of the star, whereas the values from the central part are mostly unique.

\section{Conclusion}

In this work we have investigated the spectral statistics of exponentially weighted ensembles of
full binary trees and $p$-branching star graphs. We have shown that corresponding spectral densities demonstrate peculiar ultrametric structure, typical for sparse graphs and for ensembles of linear chains exponentially distributed in lengths.

Both models, presented in the work, the fully branching 3-valent trees and the star-like graphs,
have interesting peculiarities which are worth mentioning:
\begin{itemize}
\item The spectral density of a \emph{single} tree already shares the ultrametric behavior, which is changed only by scaling factor when passing from a single tree to the ensemble of trees exponentially distributed in their sizes.
\item The spectral density of the exponential ensemble of trees looks very similar to the one of linear chains, sharing the same asymptotic behavior of tails of the distribution, $\sim e^{-c/\sqrt{\lambda_{max}- \lambda}}$, where for $p$-branching trees one has $\lambda_{max} = 2\sqrt{p-1}$ (i.e. for linear chains $p=2$ and $\lambda_{max}^{lin}=2$, while for 3-branching trees $p=3$ and $\lambda_{max}^{tree}= 2\sqrt{2}$).
\item The spectral statistics of $p$-branching star-like graphs strongly depends on $p$ (the branching of the root point) and on $\mu$ (the parameter in the exponential distribution of graphs sizes).
\end{itemize}

The edge singularity $\sim e^{-c/\sqrt{\lambda_{max}- \lambda}}$ at $\lambda\to \lambda_{max}$
reproduces the corresponding Lifshitz tail of the one-dimensional Anderson localization in a random
Schr\"odinger operator \cite{lifshitz, pastur, kirsch} with strong disorder: $\rho(E)\sim
e^{-1/E^{d/2}}$ where $E=\lambda_{max}-\lambda$ and $d=1$. The appearance of the edge singularity
$\rho(E)\sim e^{-1/\sqrt{E}}$ in our situation is purely geometric, it does not rely on any
entropy-energy-balance consideration like optimal fluctuation \cite{lifshitz,pastur}. Moreover, we
should emphasize an universality of a Lifshitz tail: typically, Anderson localization appears
for random three-diagonal operators with the randomness on the main diagonal, while in our case
there is no diagonal disorder and the edge singularity has a kind of a number theoretic signature.

In a more practical setting, our analysis demonstrates that measuring relaxational times of dilute solutions of dendrimers, or of linear chains exponentially distributed in lengths, or of sparse clusters, we arrive in all that cases at very similar hierarchically organized spectral densities. So, our study can be regarded as a warning for chemists: the relaxational spectrum is not determined by the cluster topology, but has rather the number-theoretic origin, reflecting the peculiarities of the rare-event statistics typical for one-dimensional systems with a quenched structural disorder.

Very promising looks the similarity of spectral density of individual dendrimer and of ensemble of linear chains with exponential distribution in lengths. The fact that dendrimers mimic some properties of ensembles of one-dimensional disordered systems signifies that dendrimers could be served as simple disorder-less toy models of one-dimensional systems with quenched disorder.

Investigation of spectral properties of star-like graphs with different branchings should be compared to the results of the work \cite{root}, where two different techniques have been considered simultaneously: the study of spectral properties of the graph adjacency matrix for finite graphs, and the study of singularities of the grand canonical partition function. The information acquired from the spectral density of the graph (or of ensemble of graphs) can be straightforwardly translated to the diffusion of point-like excitations on the graph (on ensemble of graphs). The interplay between the length of arms and the branching number can lead to the localization of such excitations due to purely entropic reasons.

\begin{acknowledgments}
The authors are very grateful to M. Tamm and K. Polovnikov for numerous discussions and valuable
comments. SN acknowledges the support by RFBR grant 16-02-00252 and by EU-Horizon 2020 IRSES project DIONICOS (612707). The work of VK and YM at Skolkovo Institute of Science and Technology is supported in part by the grant of the President of Russian Federation for young PhD MK-9662.2016.9 and by the RFBR grant 15-07-09121a. The work at LANL was carried out under the auspices of the National Nuclear Security Administration of the U.S. Department of Energy under Contract No. DE-AC52-06NA25396.

\end{acknowledgments}


\begin{thebibliography}{9}

\bibitem{brouwer2011spectra} A. E. Brouwer and W. H. Haemers, \emph{Spectra of graphs},
Springer Science \& Business Media (2011)

\bibitem{mohar1997some} B. Mohar, \emph{Some applications of Laplace eigenvalues of graphs},
(Springer: 1997)

\bibitem{chung1997spectral} F. R. Chung, \emph{Spectral graph theory}, (American Math. Soc.: 1997)

\bibitem{avetisov2015native} V. Avetisov, P. Krapivsky, and S. Nechaev, Native ultrametricity of
sparse random ensembles, J. Phys. A: Math. Theor., {\bf 49} 035101 (2015)

\bibitem{lifshitz} I. M. Lifshitz, Theory of fluctuation levels in disordered systems, Sov. Phys.
JETP, {\bf 26} 462 (1968)

\bibitem{planat} M. Planat, $1/f$ Frequency Noise in a Communication Receiver and the Riemann
Hypothesis, [the series Lecture Notes in Physics], Noise, Oscillators and Algebraic Randomness,
{\bf 550} 265 (2000)

\bibitem{drosoph} C. Kamp and K. Christensen, Spectral analysis of protein-protein interactions
in \emph{Drosophila melanogaster}, Phys. Rev. E {\bf 71} 041911 (2005)

\bibitem{mairal2014sparse} J. Mairal, F. Bach, and J. Ponce, Sparse Modeling for Image and Vision
Processing. Foundations and Trends in Computer Graphics and Vision, {\bf 8} 85 (2014)

\bibitem{peleg2010exploiting} T. Peleg, Y. C. Eldar, and M. Elad, Exploiting statistical
dependencies in sparse representations for signal recovery,  IEEE Trans. Signal Processing, {\bf
60}, 2286 (2012)

\bibitem{vanden2012rare} E. Vanden-Eijnden and J. Weare, Rare event simulation of small noise
diffusions, Comm. Pure Appl. Math., {\bf 65} 1770 (2012)

\bibitem{rabadan} V. Trifonov, L. Pasqualucch, R. Dalla-Favera, and R. Rabadan, Fractal-like
distributions over the rational numbers in high-throughput biological and clinical data, Scientific
Reports, {\bf 1} 191 (2011)

\bibitem{furstenberg2012analytical} F. F\"urstenberg, M. Dolgushev, and A. Blumen, Analytical model
for the dynamics of semiflexible dendritic polymers, J. Chem. Phys., {\bf 136} 154904 (2012)

\bibitem{mez} M. Mezard, G. Parisi, and M. Virasoro, \emph{Spin glass theory and beyond},
(World Scientific: 1987)

\bibitem{fra} H. Frauenfelder, Nature Str. Biol., {\bf 2} 821 (1995)

\bibitem{markelov1} Y. Gotlib, D. Markelov, Theory of the relaxation spectrum of a dendrimer macromolecule, Polym. Sci., Ser. A {\bf 44}, 1341 (2002)

\bibitem{markelov2} D. Markelov, S. Lyulin, Y. Gotlib, A. Lyulin, V. Matveev, E. Lahderanta, and A. Darinskii, Orientational mobility and relaxation spectra of dendrimers: theory and computer simulation, J. Chemi. Phys. {\bf 130}, 044907-1/9 (2009)

\bibitem{gennes} P.-G. de Gennes, \emph{Scaling concepts in polymer physics},
(Cornell Univ. Press: 1979)

\bibitem{grosberg2015statistics} A. Y. Grosberg and S. K. Nechaev, From statistics of regular
tree-like graphs to distribution function and gyration radius of branched polymers, J. Phys. A:
Math. Theor., {\bf 48} 345003 (2015)

\bibitem{rojo2005spectra} O. Rojo, R. Soto, The spectra of the adjacency matrix and Laplacian
matrix for some balanced trees, Linear algebra and its applications, Elsevier, {\bf 403} 97 (2005)

\bibitem{rojo2007explicit} O. Rojo and M. Robbiano, An explicit formula for eigenvalues of Bethe
trees and upper bounds on the largest eigenvalue of any tree, Lin. Algebra Appl., {\bf 427} 138
(2007)

\bibitem{mehta} M. Mehta, \emph{Random Matrices} (Academic Press: 2004)

\bibitem{tracy} C.A. Tracy and H. Widom, Level-spacing distributions and the Airy kernel, Comm. Math. Phys. {\bf 159}, 151 (1994)

\bibitem{pastur} I. M. Lifshitz, S. A. Gredeskul, and L. A. Pastur, \emph{Introduction to the theory of disordered systems}, (Wiley-Interscience: 1988)

\bibitem{alloys} J. Ziman, \emph{Models of disorder} (1979)

\bibitem{root} S. Nechaev, M. Tamm, O. Valba, Paths counting on simple graphs: from escape to localization, J. Stat. Mechanics, to appear (2017)

\bibitem{kirsch} W. Kirsch and I. Veselic, Lifshitz Tails for a Class of Schr\"{o}dinger Operators
with Random Breather-Type Potential, Lett. Math. Phys. {\bf 94} 27 (2010)



\end{thebibliography}
\end{document}